\documentclass[twocolumn]{aastex631}
\usepackage{amsmath}

\usepackage{float}

\begin{document}

\title{Parkes transient events: II. Pulsar single pulses database containing raw data segment}

\correspondingauthor{Xuan Yang, Songbo Zhang}
\email{yangxuan@pmo.ac.cn, sbzhang@pmo.ac.cn}

\author{Xuan Yang}
\affiliation{Purple Mountain Observatory, Chinese Academy of Sciences, Nanjing 210023, China}
\affiliation{School of Astronomy and Space Sciences, University of Science and Technology of China, Hefei 230026, China}

\author{Songbo Zhang}
\affiliation{Purple Mountain Observatory, Chinese Academy of Sciences, Nanjing 210023, China}
\affiliation{CSIRO Space and Astronomy, Australia Telescope National Facility, PO Box 76, Epping, NSW 1710, Australia}

\author{Le-Yu Tang}
\affiliation{Nanjing Foreign Language School, Nanjing 210008, China}

\author{L. Toomey}
\affiliation{CSIRO Space and Astronomy, Australia Telescope National Facility, PO Box 76, Epping, NSW 1710, Australia}

\author{Xuefeng Wu}
\affiliation{Purple Mountain Observatory, Chinese Academy of Sciences, Nanjing 210023, China}
\affiliation{School of Astronomy and Space Sciences, University of Science and Technology of China, Hefei 230026, China}


\begin{abstract} 
We have re-processed single pulse candidates from the first four years (1997$-$2001) of the Parkes Multibeam receiver system observations, creating a new Parkes transient database (PTD II) that contains 165,592 single pulses from 363 known pulsars. 
Unlike previous databases, PTD II preserves the critical raw data segments of each detected pulse, enabling detailed analyses of emission physics while maintaining a compact size of only 1.5 GB. 
The database employs a \emph{\sc sqlite3} structure organising pulsar metadata, observation files, and pulse events with their raw data stored in binary format. We provide processing tools for extracting and analysing single-pulse data, enabling fluence fitting and statistical analysis. Our pulsars exhibit diverse fluence distributions, such as log-normal, Gaussian, and unimodal.
Temporal analyses reveal significant evolution in emission characteristics of several pulsars, including event rate variations spanning two orders of magnitude in PSR J1602$-$5100 and PSR J0942$-$5552.
Beyond supporting fundamental pulsar emission studies, the database serves as a valuable training resource for developing new single-pulse detection algorithms in the presence of radio frequency interference. 
\end{abstract}


\section{Introduction} \label{sec:intro}


Pulsars are rapidly rotating neutron stars emitting beams of electromagnetic radiation, first discovered through their individual radio pulses, which present a perfect periodicity~\citep{first_pulsar}.
The subsequent detection of millisecond pulsars (MSPs) by~\citep{Backer82_msPSR} unveiled a new population of neutron stars exhibiting  even more extreme rotational stability.
These discoveries established pulsars as precision tools for probing extreme physics, from nuclear matter equations of state to gravitational wave detection~\citep[][and references therein]{handbook}.
Contemporary pulsar research predominantly studies averaged pulse profiles to achieve the timing precision required for these applications.

Early pulsar observational studies in the late 1960s$-$1980s also revealed that pulsar individual pulses could uncover fundamental emission phenomena: ~\citet{Drake68_drift} identified drifting subpulses exhibiting systematic phase variations; ~\citet{Nulling} characterised nulling episodes where emission abruptly ceases; ~\citet{Bartel82_modeChange} documented mode-changing behavior involving discrete profile alterations; and ~\citet{crab} detected giant pulses with intensities exceeding the mean by orders of magnitude.
The discovery of Rotating Radio Transients (RRATs) by~\citet{rrat} further underscored the critical importance of single-pulse studies, as these sources exhibit sporadic emission detectable more easily through single-pulse searches rather than periodicity analyses~\citep{burke_singlepulsesearch}.
Therefore, single-pulse investigations remain essential for understanding both canonical pulsar emission physics and exotic neutron star populations.

Because time averaged pulse profiles of most radio pulsars are typically stable, pulsar research has traditionally focused on profiles obtained by folding many rotational periods. Even studies of well-known single-pulse phenomena, such as drifting, mode changing, and nulling, have largely emphasized population-level comparisons or focused on rare events like giant pulses. However, the single-pulse emission of many sources is highly variable and complex. Recent observations of RRAT J1913+1330 have also revealed dramatic pulse-to-pulse variations in profile shape, width, peak flux, and fluence, with energy distributions spanning up to three orders of magnitude, reminiscent of those seen in repeating FRBs~\citep{j1913}. Large, homogeneous single-pulse datasets, such as the one presented in this study, are therefore essential for probing the radiation physics underlying all short-duration coherent radio pulses from neutron stars, including those from RRATs, magnetars, and FRBs.

With the catalogued pulsar population now approaching 4,000~\footnote{\url{https://www.atnf.csiro.au/research/pulsar/psrcat/}}~\citep{atnfcatlog}, the computational challenges of single-pulse studies using raw search-mode datasets have grown exponentially.
In the face of the escalating data volumes and processing complexity, systematic databases become essential for population-level investigations.

In the first paper of this series, ~\citet{songbodatabase} compiled a reference catalogue of Parkes transients~(Parkes Transient Database, PTD I)~\footnote{\url{https://doi.org/10.25919/5e33a52c18a17}}, which standardised the storage and access of single-pulse detections from four years of Parkes Multibeam observations containing 568 million candidates.
This database serves as a resource for exploring radio emission phenomena and searching for transient events like RRATs and Fast Radio Bursts (FRBs). 
PTD I primarily focused on presenting the data processing pipeline and identifying unknown radio transients. 
By providing access to single-pulse observations, it enables statistical studies of radio transients detected by Parkes~\citep{Zhang24_onefOff} and machine learning applications~\citep{Yang21}.

%

In Section~\ref{sec:obs}, we provide a description of the data and processing pipeline used in this study. The database structure and single pulse analysis example are presented in Section~\ref{sec:Results}, while Section~\ref{sec:discuss} includes a discussion of the potential application of our database.

\section{Dataset and single pulse extraction} \label{sec:obs}
We now present PTD II~\footnote{\url{https://doi.org/10.25919/34am-zx04}} which focuses on a subset of pulses from known pulsars extracted from PTD I. The raw data segments of these pulses are also stored in PTD II.

The PTD I contains the searching results of all the observations during the first four years (from 1997 to 2001) of operation of the Parkes Multibeam receiver system~\citep{multibeam}. The data sets used in creating the PTD I were all obtained with the primary goal of discovering pulsars.
All data during the four years were processed with a single-pulse search pipeline using \emph{\sc PRESTO}~\footnote{\url{http://www.cv.nrao.edu/~sransom/presto/}}. The PTD I~\citep{songbodatabase} has recorded 568,736,756 pulse candidates generated from the pipeline~\citep[see][for more details]{songbodatabase}.


Our pipeline in PTD II makes use of \emph{\sc sqlite3}~\footnote{\url{https://www.sqlite.org/}} to automatically select single-pulse candidates from the file segments in PTD I. Only candidates that meet both following criteria are confirmed as single-pulse detections from a pulsar:\\
\begin{enumerate}
\item $\rm D_{\rm s}<0.23 \rm \,deg$ ,\label{dist}
\item $\left|\rm DM_{detection}-\rm DM_{pulsar}\right|<15\,\rm cm^{-3}\, \rm pc$ , 
\item $\rm SNR>8$ ,
\end{enumerate}
where $D_s$ is the distance between the location of the pulsar and the pointing centre of the beam, $\rm DM_{detection}$ is the dispersion measure given by the single pulse searching pipeline, $\rm DM_{pulsar}$ is the dispersion measure of the pulsar obtained from the ATNF Pulsar Catalogue, $\rm SNR$ is the signal-to-noise ratio given by \emph{\sc PRESTO}. We adopt the $0.23 \rm \,deg$ which is the full width at half power~(FWHP) beamwidth of the central beam of the multibeam system~\footnote{\url{https://www.parkes.atnf.csiro.au/research/multibeam/.overview.html}}. The number $15\,\rm cm^{-3}\, \rm pc$ is the average DM searching step size of in PTD I.

The filtered files are now confirmed to contain a pulse or a train of pulses.
We then extract the pulses from the filtered files, and create a new small psrfits file~\footnote{\url{https://www.atnf.csiro.au/research/pulsar/psrfits_definition/Psrfits.html}} segment that contains the single pulse detection. 
The \emph{\sc python} package \emph{\sc astropy.io.fits} was used to extract the time samples and modify the header. The NAXIS2, MJD and NSBLK in the header of the file segment are updated, where NAXIS2 is the number of rows in the data table~(number of sub-integration), MJD is the Modified Julian Date of the first time sample in the file, NSBLK is the number of samples per row. The information of the single pulse and the file segment is then stored in PTD II.

\section{Parkes Transient Database II}
\label{sec:Results}

\subsection{Database configuration}
\begin{figure*}
    \begin{center}
    \includegraphics[width=0.9\textwidth]{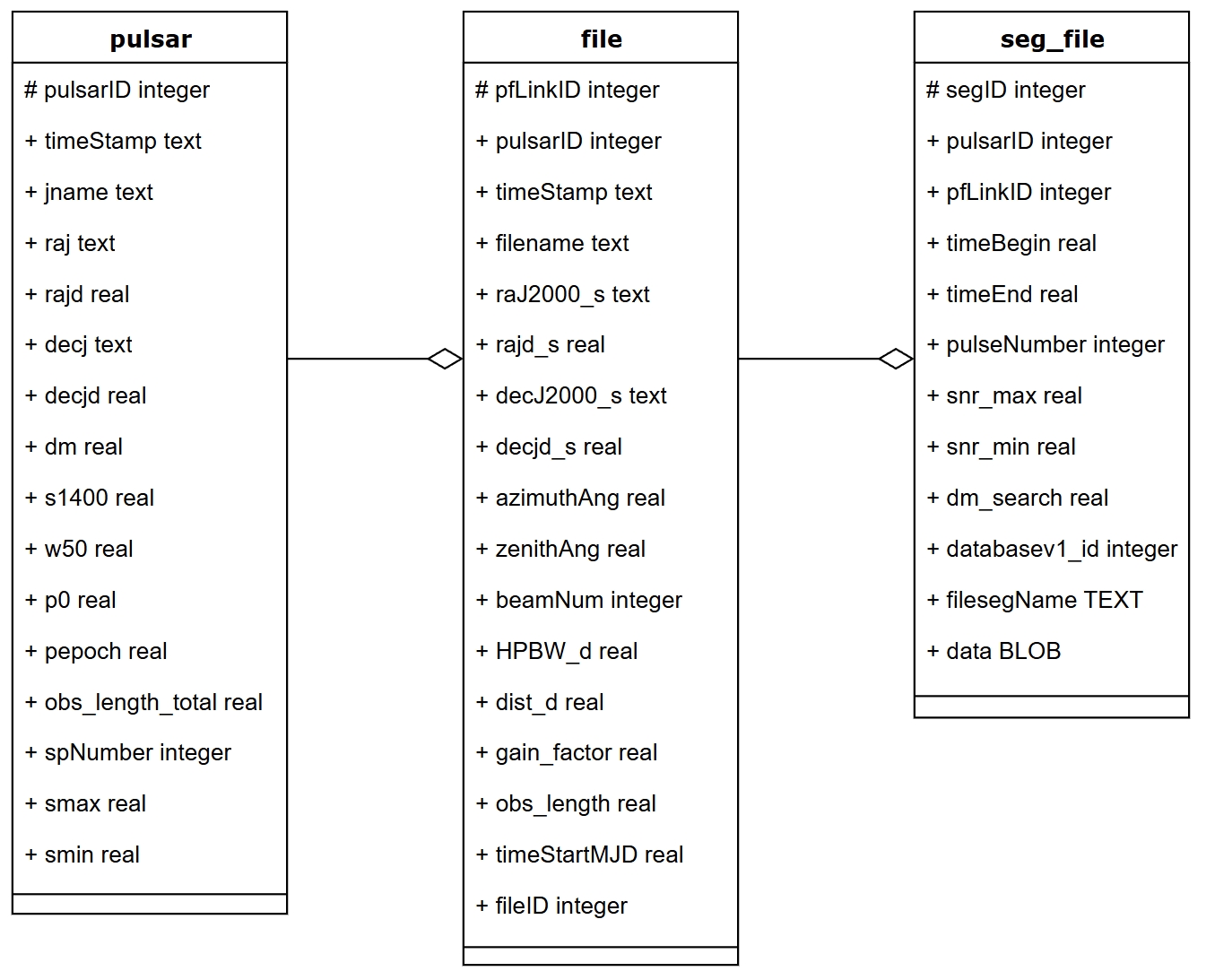}
    \caption{Chart of the database schema.}
    \label{figure:database}
    \end{center}
\end{figure*}

We have created a \emph{\sc sqlite3} database to store the pulsar information and raw data of the pulsar single pulses. The database schema and description of the parameters are listed in Table \ref{table:parameter} and graphically displayed in Figure~\ref{figure:database}.

In brief, each pulsar that has single pulses is identified by the \emph{\sc pulsar} table. This table contains information such as the DM and period based on the ATNF Pulsar Catalogue v2.6.3~\footnote{\url{https://www.atnf.csiro.au/research/pulsar/psrcat/}}~\citep{atnfcatlog}.  
The obs\_length\_total shows the integration time of the pulsar, where all the observation files with beam pointing distances within 0.23 deg of the pulsar position are included~(no matter single pulses detected or not). The spNumber shows the total number of single pulses detected.

We have linked specific files to known pulsars using the \emph{\sc file} table. 
These files contain single-pulse detections from the pulsars. 
Observation parameters 
such as filename, pointing direction, angular separation from the pulsar,  
and start time (MJD) are stored in the \emph{\sc file} table.

Each file is connected to multiple \emph{\sc fileSegment} tables. Most file segments contain a single-pulse detection. For some pulsars, the delay time derived from equation~\ref{eq:dmtime} describing the frequency dependent time of arrival introduced by the dispersion approaches the pulsar's period, resulting in single file segments containing multiple pulses. 
\begin{equation}
    \Delta t_{\rm DM}=4.15{\rm ms}\times\left(\frac{\nu_{\rm l}^{-2}-\nu_{\rm h}^{-2}}{\rm GHz}\right)\times\left(\frac{\rm DM}{\rm cm^{-3}\,\rm pc}\right).
	\label{eq:dmtime}
\end{equation}
Here $\nu_{\rm l}$ and $\nu_{\rm h}$ epresent the lower and upper frequency bounds. 
The \emph{\sc fileSegment} table stores information such as  the start time of the first pulse in the pulse train, the end time of the last pulse in the pulse train, DM, S/N, pulse number, 
and raw data of the single pulse 
in binary format.

\begin{table*}[htbp]
  \begin{scriptsize}
  \caption{The parameters stored in the database.\label{table:parameter}}
  \renewcommand\arraystretch{1.0}
  \setlength{\tabcolsep}{2.0mm}  
  \begin{center}
  \begin{tabular}{@{\extracolsep{\fill}}lcc}
  \hline
 Name  &    Type      &  Description   \\
\hline
\hline
\textbf{pulsar} &   &  pulsars parameters provided by the latest ATNF pulsar catalogue \\ 
pulsarID & integer  &  The primary key \\ 
timeStamp & text & Time stamp of insertion\\
jname &  text & Pulsar name based on J2000 coordinates\\
raj &  text & Right ascension (J2000) (hh:mm:ss.s)\\
rajd &  real & Right ascension (J2000) (deg)\\
decj &  text & Declination (J2000) (dd:mm:ss)\\
decjd &  real & Declination (J2000) (deg)\\
 dm &  real & Dispersion measure (${\rm cm}^{-3}{\rm pc}$)\\
s1400 &  real & Mean flux density at 1400 MHz (mJy)\\
w50 &  real & Width of pulse at 50\% of peak (ms)\\
p0 &  real & Barycentric period of the pulsar (s)\\
pepoch &  real & Epoch of period~(MJD)\\
obs\_length\_total &  real & Total integration time of the Pulsar (s) \\
spNumber &  integer &  Number of single pulses dectected\\
smax &  real & Max S/N of the single pulses\\
smin &  real & Min S/N of the single pulses\\
\hline
\textbf{file} &   &  A file contains the observation from one beam \\
pfLinkID &  integer & The primary key\\
pulsarID &  integer & Identifier for the related pulsar\\
timeStamp &  date & Time stamp of insertion\\
filename &  text & Name of the file\\
raJ2000\_s &  text & Right ascension in J2000 coordinates of the pointing centre of the beam (hh:mm:ss.ssss)\\
rajd\_s &  real & Right ascension in J2000 coordinates of the pointing centre of the beam (deg)\\
decJ2000\_s &  text & Declination in J2000 coordinates of the pointing centre of the beam (dd:mm:ss.sss)\\
decjd\_s &  real & Declination in J2000 coordinates of the pointing centre of the beam (deg)\\
azimuthAng &  real & Azimuth angle (deg)\\
zenithAng &  real & Zenith angle (deg)\\
beamNum &  integer & Beam number for multibeam systems (1=central beam)\\
HPBW\_d &  real & Half power beamwidth of the beam (deg)\\
dist\_d &  real & Distance between the location of the pulsar and the pointing centre of the beam (deg)\\
 gain\_factor & real& The gain degrading factor caused by the dist\_d, gain\_factor=1 when the pulsar is located at the centre of the beam\\
obs\_length &  real & The full duration of the file (s)\\
timeStartMJD &  real & MJD of the first sample for the observation in UTC time\\
fileID &  integer & fileID from database version I\\
\hline
\textbf{seg\_file} &   &  A file segment contains the grouped candidates in one group \\
segID &  integer &  The primary key\\
pfLinkID  & integer &   Identifier for the searched file, link to the table of file\\
timeBegin  & real &  time of the segment begin\\
timeEnd  & real &  time of the segment end\\
pulseNumber  & integer &   Number of single pulses in the segment\\
snr\_max  & real &  Max S/N of the single pulses\\
snr\_min  & real &  Min S/N of the single pulses\\
dm\_search  & real &  DM where the candidate detected with maximum S/N (${\rm cm}^{-3}{\rm pc}$)\\
databasev1\_id  & integer & fileSegmentID from database version I\\
filesegName  & text &  Name of the file segment\\
data  & blob &  The file segment stored in binary form\\
\hline
\hline
\end{tabular}
\end{center}
\end{scriptsize}
\end{table*}

\subsection{Pulsar single pulse data in database}
The final database contains 165,592 single pulses from 363 pulsars, with 157,517 file segments. Among these, 3,383 segments contain multiple pulses. 
Table~\ref{table:properties} catalogues the single pulse detection statistics for each pulsar.

\begin{figure*}
    \begin{center}
    \includegraphics[width=0.9\textwidth]{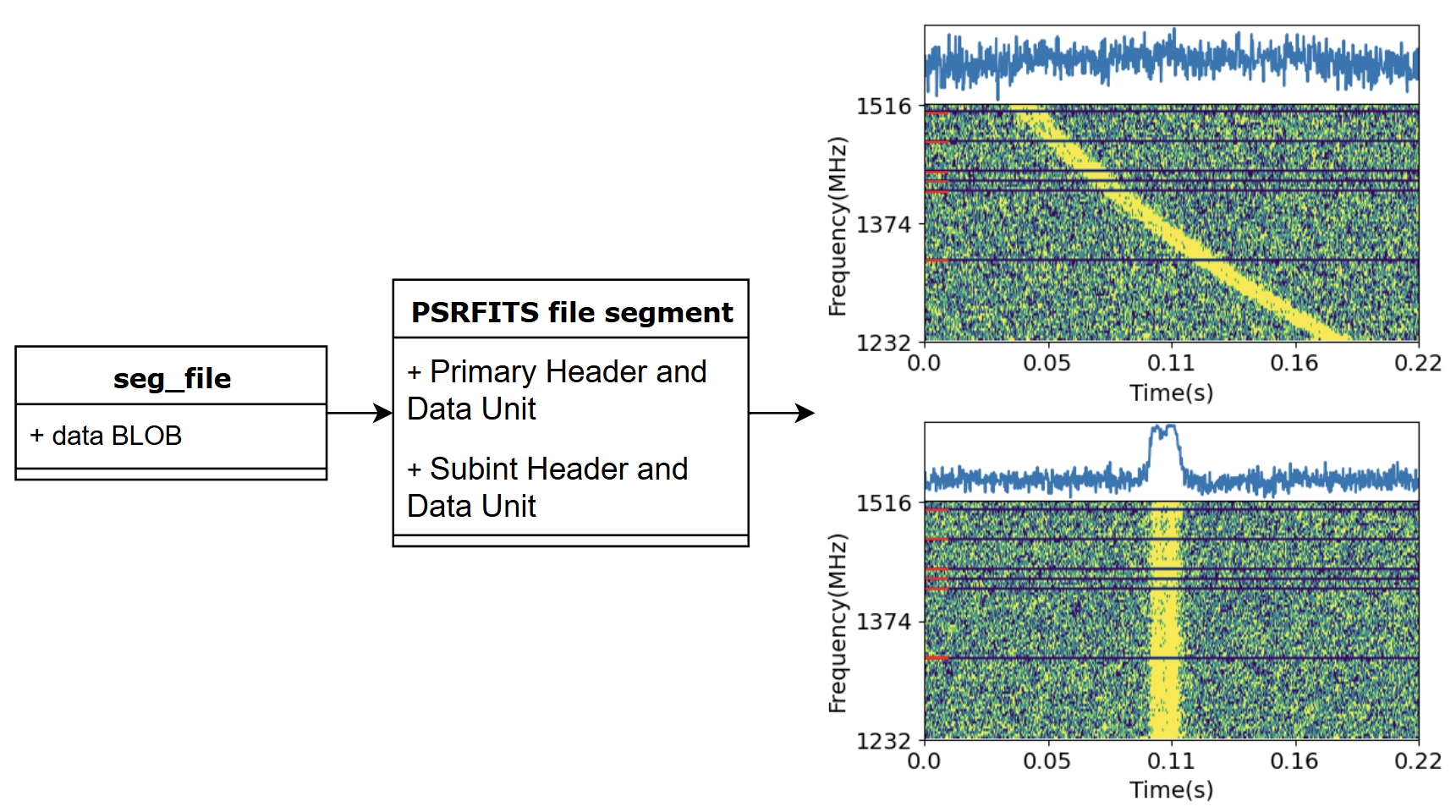}
    \caption{The process executed by the script \emph{\sc get\_pulsar\_pub.py}: it first extracts the blob data from the PTD II (left), subsequently saves this data as a PSRFITS file (middle), and finally generates a plot of the single pulse using the PSRFITS file (right). For the right panel, from top to bottom are: integrated pulse profile using an arbitrary flux density scale, time–frequency plane, integrated pulse profile after being de-dispersed at the pulsar DM, single pulse spectrum after being de-dispersed at the pulsar DM. The blue narrow lines with short red lines on the left side are the marked narrowband RFIs.}
    \label{figure:extract}
    \end{center}
\end{figure*}

We provide the \emph{\sc get\_pulsar\_pub.py} software tool to extract single-pulse data and analyse fluence distributions for a specific pulsar. 
For example, use the command:

\begin{gather*}
    \text{python get\_pulsar\_pub.py -db Pulsar\_fits\_database.db} \\
    \text{-j J1745-3040,}
\end{gather*}

all 3,516 single pulse file segments from PSR J1745$-$3040 will concurrently be extracted as psrfits format and
saved to disk.
Figure~\ref{figure:extract} shows the workflow:
blob~\footnote{The binary large object, a data type used to store large amounts of unstructured binary data} data is extracted from PTD II, saved as an individual PSRFITS file which can be used to generate dynamic spectra.

\emph{\sc get\_pulsar\_pub.py} could also obtain the pulsar fluence analysis. 
Flux density for an event is estimated using:
\begin{equation}
    S=\frac{\sigma \  S/N \ T_{\rm sys}}{G \sqrt{\Delta \nu  N_p t_{\rm obs}}},
	\label{eq:slim}
\end{equation}
where $\sigma=1.5$ is a loss factor due to 1-bit digitisation~\citep{Manchester2001}, $T_{\rm sys}$ is the system temperature, $G$ is the telescope antenna gain\footnote{\url{https://www.parkes.atnf.csiro.au/research/multibeam/.overview.html}}, $\Delta \nu$ is the observing bandwidth, $N_p$ is the number of polarization channels, and $t_{\rm obs}$ is the integration time. Considering the distance between the location of the pulsar and the pointing centre of the beam, the $\rm G$ used in equation~\ref{eq:slim} is multiplied by a degrading factor $G_f$, which is stored in the \emph{\sc file} table. The $G_f$ is estimated from the multibeam gain model published by \cite{multibeam_gain}  and takes into account the position of the pulsar within the beam.
Fluence is then obtained by integrating $S$ over the burst duration. 

In \emph{\sc get\_pulsar\_pub.py}, the fluence estimation will be carried out after a narrowband radio frequency interference~(RFI) mitigation step, the channels exceeding the mean value by three standard deviations will be marked and removed, and the bandwidth for fluence calculation will be updated to the new bandwidth. The software will also create a waterfall plot for each single pulse after estimating the fluence, as shown in the right panel of figure~\ref{figure:extract}. The top half of the plot shows the time–frequency plane and the integrated pulse profile before de-dispersing, the lower half shows the time–frequency plane and the integrated pulse profile after de-dispersing using the DM of the pulsar.

\begin{figure*}
    \begin{center}
    \includegraphics[width=0.9\textwidth]{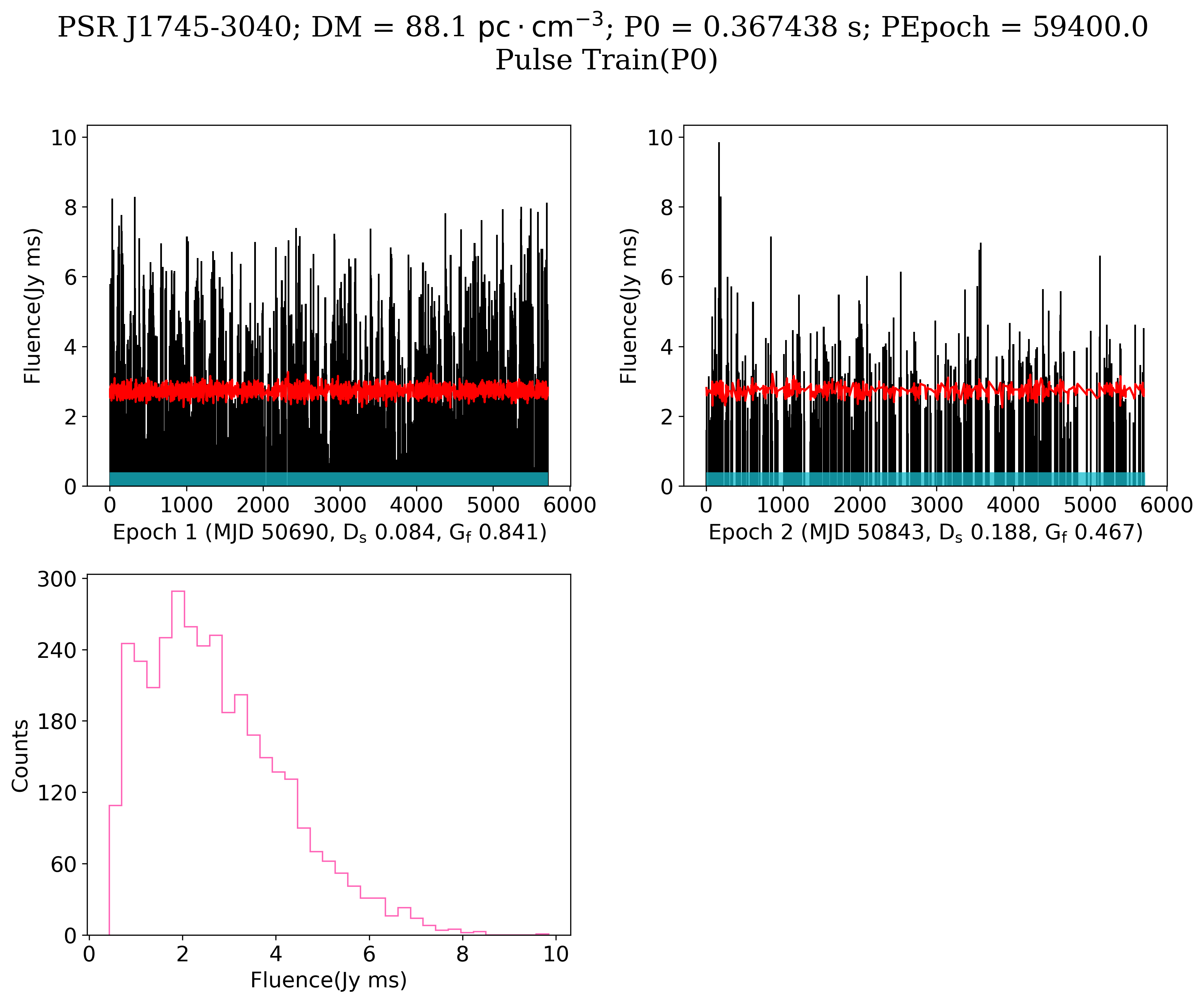}
    \caption{Fluence of 3,516 single pulses of PSR J1745$-$3040 from 2 observation epochs. The X-axis represents the pulse train number relative to the start of the observation, measured in units of the rotating period (P0). Each observation epoch is indicated by a blue background. The red solid lines represent the RMS level variation scaling to mean fluence level.}
    \label{figure:J1745}
    \end{center}
\end{figure*}

By typing:
\begin{gather*}
    \text{python get\_pulsar\_pub.py -db Pulsar\_fits\_database.db} \\
    \text{-fluence -j J1745-3040,}
\end{gather*}
the fluences of single pulses from PSR J1745$-$3040 will be estimated, the result is shown in Figure~\ref{figure:J1745}. The software will also generate a text file which contains the data used to plot the figure. In total 3,516 single pulses (black bars) were detected, with normalised relative fluence coefficient (NRFC, red solid lines) overlaid:  
\begin{equation}
    \rm NRFC=\frac{\rm RMS(t)}{\rm{RMS_{ mean}}}\times \rm{\rm fluence_{mean}},
	\label{eq:nrfc}
\end{equation}
where the NRFC quantifies baseline Root Mean Square~(RMS) variations relative to the mean fluence, the $\rm RMS(t)$ is the RMS value at the time when the corresponding single pulse arrives, the $\rm{RMS_{ mean}}$ is the average RMS value of $\rm RMS(t)$. Stable noise levels across epochs are evident from the red curves, while blue backgrounds indicate telescope integration times. The bottom subplot shows the fluence distribution of all detected single pulses.


\begin{figure}[htbp]
    \begin{center}
    \includegraphics[width=0.45\textwidth]{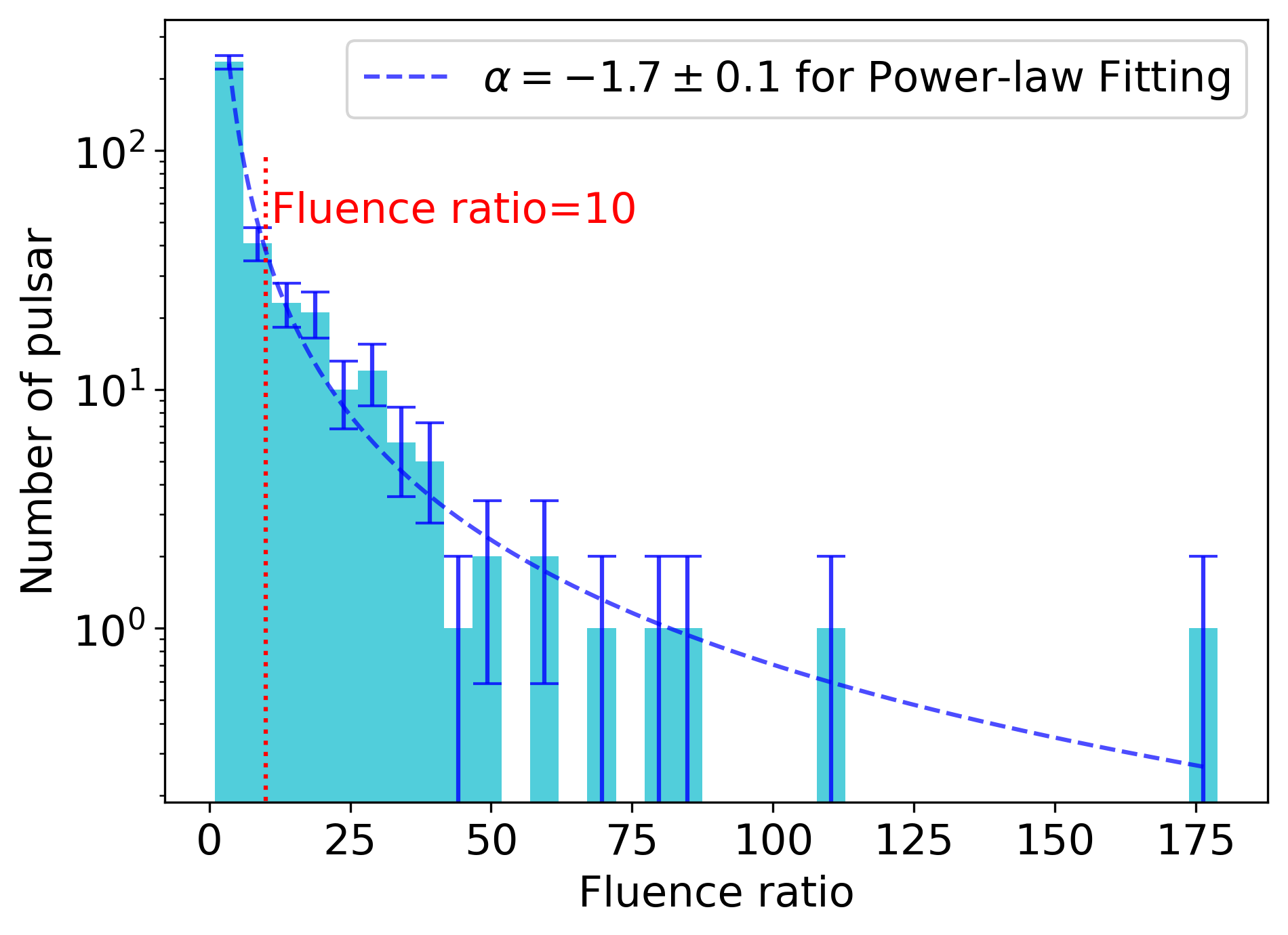}
    \caption{The fluence ratio distribution of the 363 pulsars.}
    \label{figure:fluratio}
    \end{center}
\end{figure}

The fluence ratio distribution of the 363 pulsars is shown in figure~\ref{figure:fluratio}. The blue bars indicate the pulsar counts per fluence ratio bin, while the red dashed line marks the fluence ratio $=10$. This ratio is noteworthy since broad energy distributions are atypically for regular pulsar single pulses, the intensity variations of more than one order were always attributed to the phenomenon of "giant pulses"~\citep{giantpulse}.
A total of 92~(25.3\%) of the pulsars have fluence ratios greater than 10.  
The fluence ratio distribution can be fitted by a power-law function and shown in blue dotted line, the power-law index is $\alpha=-1.7\pm 0.1$.


\section{Discussion}
\label{sec:discuss}
\subsection{Temporal variability in pulsar single-pulse emission}

The primary goal of this work was to provide a basic tool for researchers to study pulsar single pulses. All raw data files from the four-year observations were pre-processed and stored efficiently, resulting in a compact 1.5 GB database containing all search results and raw data segments. 
The database is publicly accessible, enabling users to conduct further analyses on standard computing systems.
The usage of the database follows the standard \emph{\sc sqlite3} syntax, we provide a \emph{\sc python} script to rapidly extract the file segment and fluence distribution of each pulsar. 

Our analysis of 363 pulsars identified 98 sources with over 100 single-pulse detections, sufficient to study their fluence distributions.
%
We used the log-normal function and the Gaussian function to fit the fluence distributions. The goodness-of-fit was quantified by a $\chi^2$ analysis following the same process as in \cite{single1}. The goodness-of-fit probability~($P$) was then calculated from the fit’s $\chi^2$ value. Probabilities were calculated for both the log-normal function~($P_l$) and Gaussian function~($P_g$). We consider fits with $P > 0.05$ statistically significant.

Among the 98 pulsars with $>100$ single-pulse detections, our distribution analysis reveals: (1) 25 pulsars whose fluence distribution can be fitted by a log-normal function (L). (2) 5 pulsars whose fluence distribution can be fitted using a Gaussian function (G). (3) 19 pulsars that can be fitted by both a log-normal function and a Gaussian function (U). (4) 49 pulsars that show complex distributions incompatible with either model (O).
Full fitting classification of the fluence distribution fits is provided in Table~\ref{table:properties}.

Temporal variability in single pulses has been observed in many pulsars. In the following subsection, we present four examples in detail. These four pulsars have more than 100 single-pulse detections and exhibit obvious changes in different observing epochs (average fluence or event rate).



\subsubsection{PSR J1602$-$5100}
\begin{figure*}
    \begin{center}
    \includegraphics[width=0.9\textwidth]{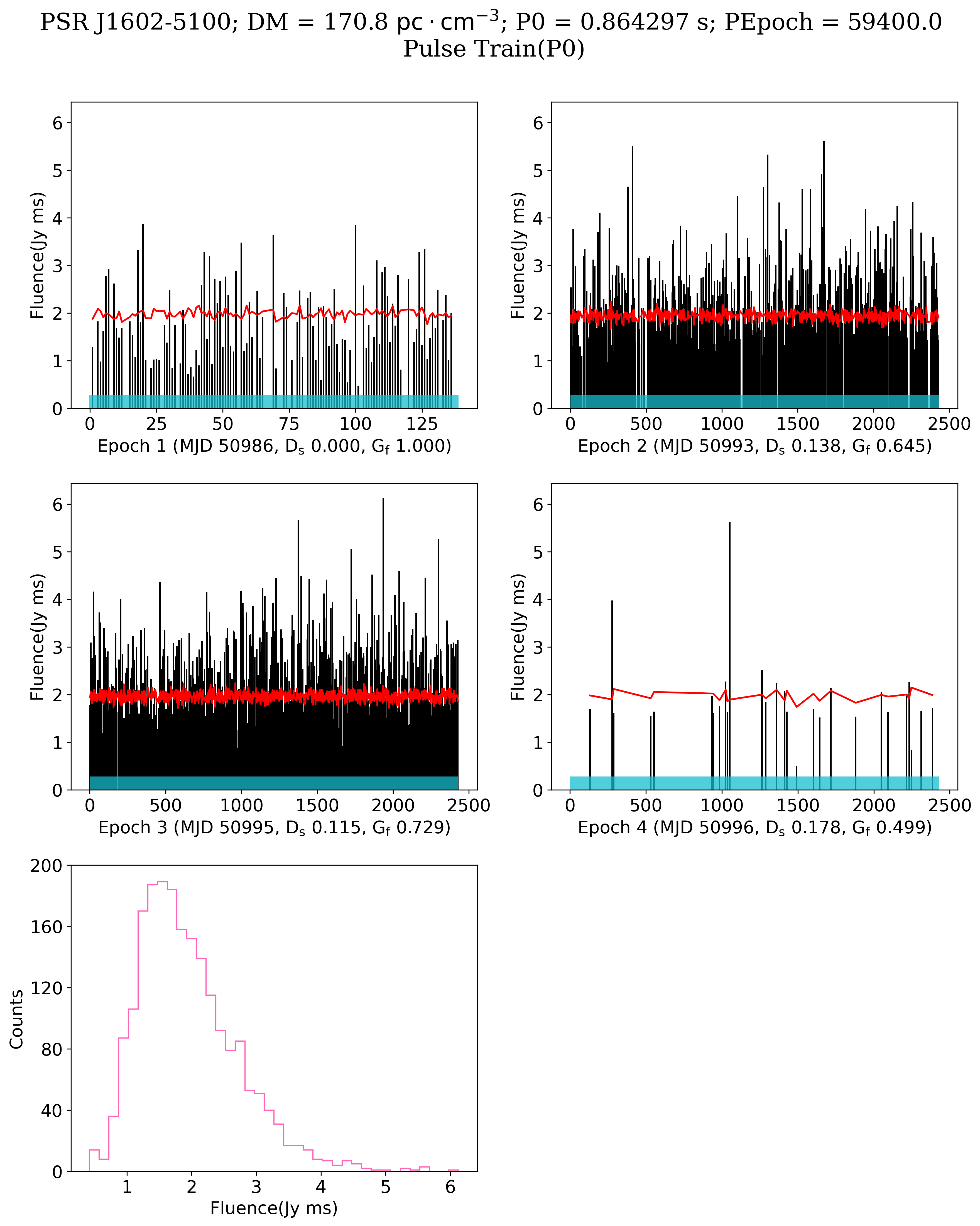}
    \caption{Fluence of 2,066 single pulses of PSR J1602$-$5100 from 4 observation epochs. The X-axis represents the pulse train number relative to the start of the observation, measured in units of the rotating period (P0). Each observation epoch is indicated by a blue background. The red solid lines represent the RMS level variation scaling to mean fluence level.}
    \label{figure:J1602}
    \end{center}
\end{figure*}

Table~\ref{table:J1602} and figure~\ref{figure:J1602} detail the variations in radiation parameters of 2,066 single pulses from J1602$-$5100 across four observational epochs (MJD 50986$-$50996). 
The pulsar exhibits significant emission evolution over 10 days, with event rates ($\rm R_{events}$) varying by two orders of magnitude. 
Although the epoch with the lowest $\rm R_{events}$ corresponds to the largest $D_s$, all four observation pointings were well within the FWHP beamwidth relative to the pulsar position. The gain degrading factor $G_f$ for each $D_s$ is also shown in the figure~\ref{figure:J1602}.
It is notable that substantial changes in the shape of the pulse profile and its correlated changes with rotation have been reported in J1602$-$5100~\citep{J16025100}.  
The variations in the event rate may be related to this emission-rotation correlation. Further dedicated observations of J1602$-$5100 are strongly encouraged to better understand the relationship between event rate fluctuations and profile shape changes.
\begin{table}[H]
\begin{center}

\caption{The emission characteristic of different epochs of J1602-5100. The fluence is shown in $\rm Mean \pm Sigma$.\label{table:J1602}}
\renewcommand\arraystretch{1.0}    
\begin{tabular}{ccccc}
\hline
MJD  &     Fluence        &     $\rm N_{events}$  	 &     	$ \rm R_{events}$   & $D_s$ 
\\
  &    (Jy ms)   &   &  $ \rm (events~period^{-1})$   &  (deg)  
\\\hline
50986 & $1.8 \pm 0.8$ & 113 & 0.815  & 0.000\\\hline
50993 & $2.0 \pm 0.7$ & 778 & 0.320  & 0.138\\\hline
50995 & $1.9 \pm 0.8$ & 1147 & 0.472  & 0.115\\\hline
50996 & $2.0 \pm 0.9$ & 28 & 0.012  & 0.178\\\hline
\end{tabular}

\end{center}
\end{table}

\subsubsection{PSR J1539$-$5626}
\begin{figure*}
    \begin{center}
    \includegraphics[width=0.75\textwidth]{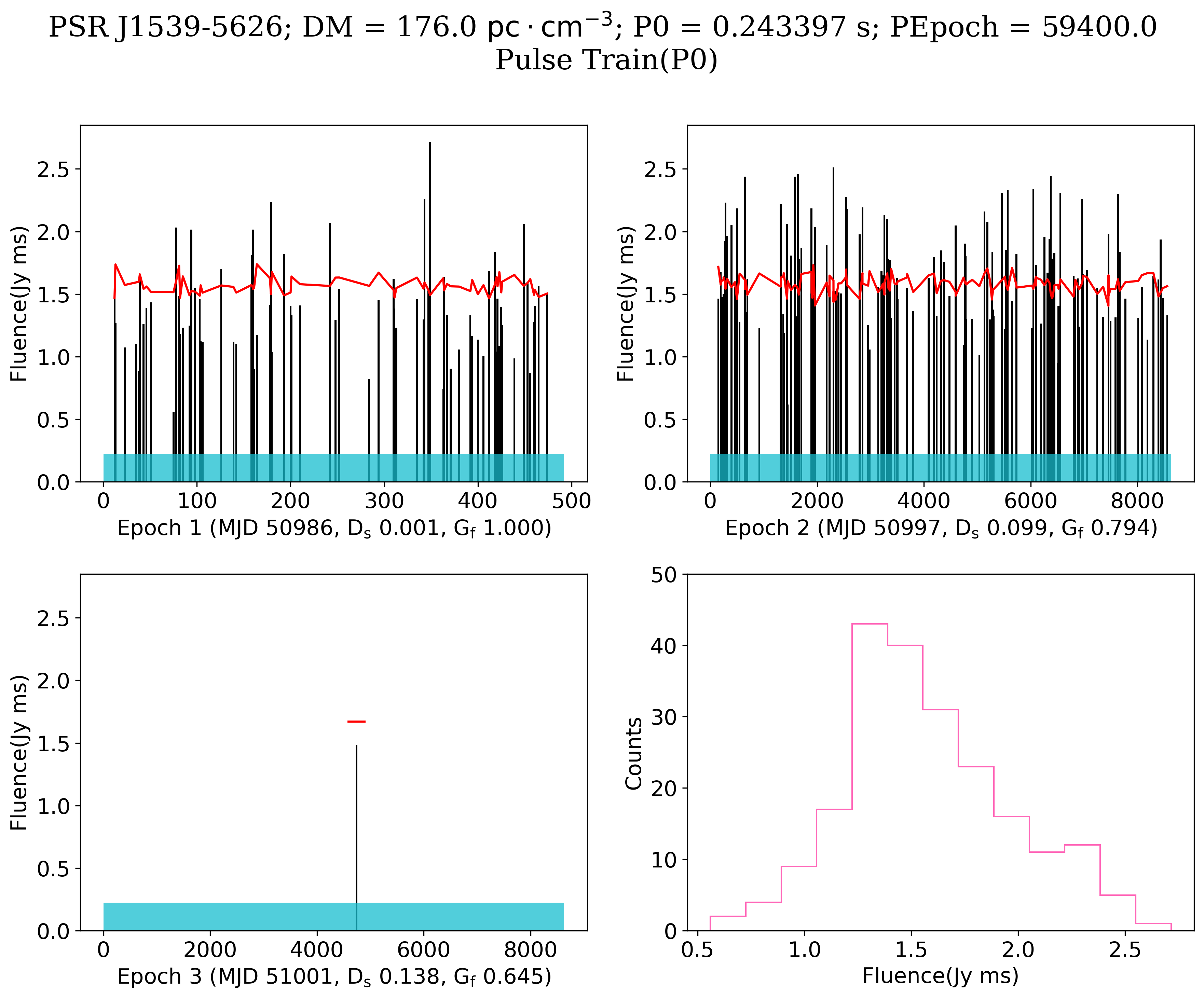}
    \caption{Fluence of 214 single pulses of PSR J1539$-$5626 from 3 observation epochs. The X-axis represents the pulse train number relative to the start of the observation, measured in units of the rotating period (P0). Each observation epoch is indicated by a blue background. The red solid lines represent the RMS level variation scaling to mean fluence level.}
    \label{figure:J1539}
    \end{center}
\end{figure*}

The emission characteristic of J1539-5626 is shown in Table~\ref{table:J1539} and figure~\ref{figure:J1539}. While fluence remains relatively stable across three epochs, the event rate varies by three orders of magnitude, decreasing from 0.146 to 0.0001 events per period.

\begin{table}[H]
\begin{center}

\caption{The emission characteristic of different epochs of J1539-5626. The fluence is shown in $\rm Mean \pm Sigma$.\label{table:J1539}}
\renewcommand\arraystretch{1.0}    
\begin{tabular}{ccccc}
\hline
MJD  &     Fluence        &     $\rm N_{events}$  	 &     	$ \rm R_{events}$   & $D_s$ 
\\
  &    (Jy ms)   &   &  $ \rm (events~period^{-1})$   &  (deg)  
\\\hline
50986 & $1.4 \pm 0.4$ & 72 & 0.146  & 0.001\\\hline
50997 & $1.7 \pm 0.4$ & 141 & 0.016  & 0.099\\\hline
51001 & $1.5 \pm 0.0$ & 1 & 0.0001  & 0.138\\\hline
\end{tabular}

\end{center}
\end{table}

\subsubsection{PSR J0942$-$5552}
\begin{figure*}
    \begin{center}
    \includegraphics[width=0.9\textwidth]{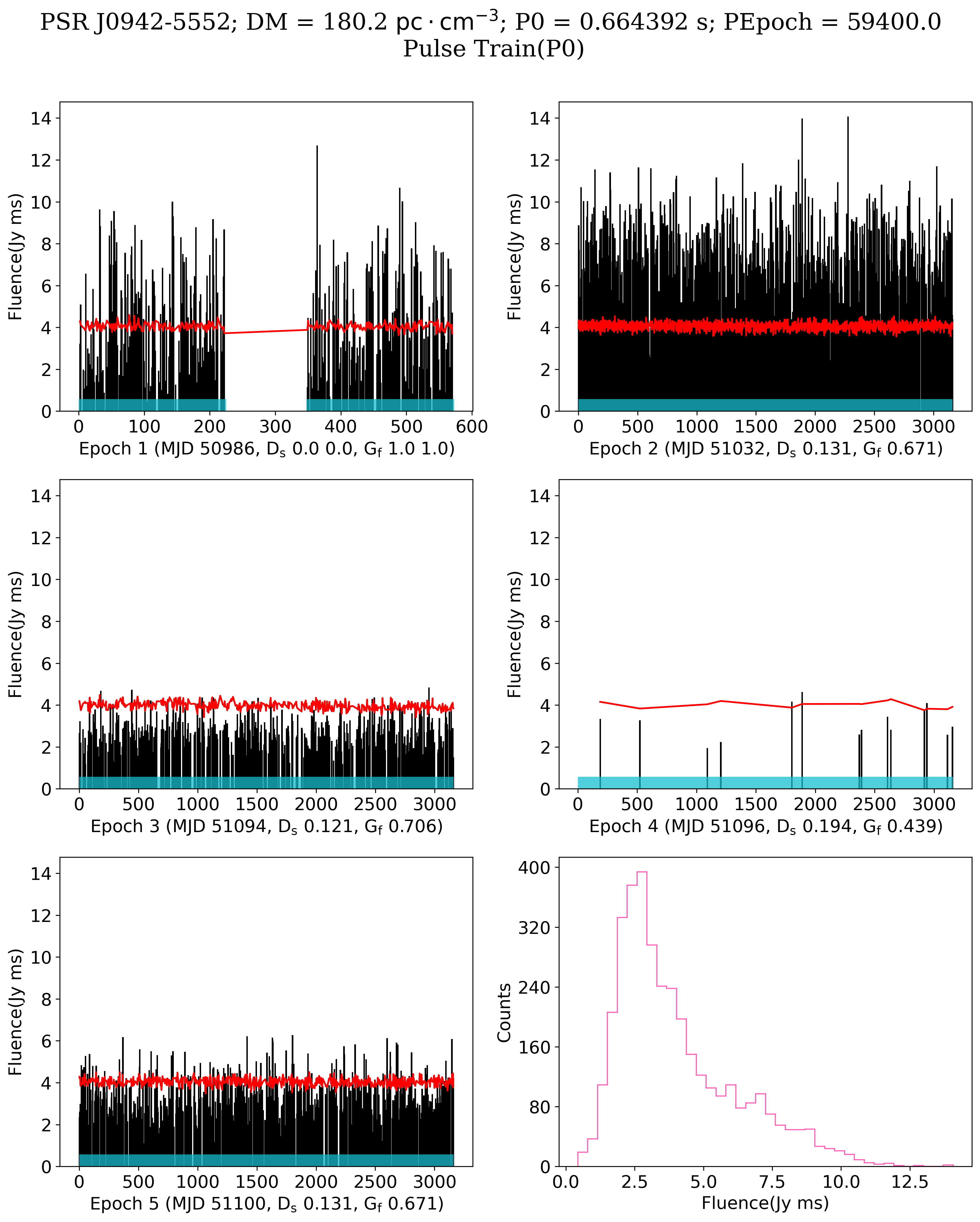}
    \caption{Fluence of 3,672 single pulses of PSR J0942$-$5552 from 5 observation epochs. The X-axis represents the pulse train number relative to the start of the observation, measured in units of the rotating period (P0). Each observation epoch is indicated by a blue background. The red solid lines represent the RMS level variation scaling to mean fluence level.}
    \label{figure:J0942}
    \end{center}
\end{figure*}

PSR J0942$-$5552 exhibits complex time-variable emission (Table~\ref{table:rateJ0942} and Figure~\ref{figure:J0942}).
The data clearly show a decline in the pulsar’s mean fluence following the first two observational epochs. During MJD 50986 and 51032, the mean fluences were measured at $4.2\pm2.4$ and $4.8\pm1.6$ Jy ms, respectively. In the subsequent three epochs (MJD 51094, 51096, and 51100), these values dropped, registering $2.6\pm0.7$, $3.2\pm0.7$, and $3.1\pm1.0$ Jy ms. 
It is notable that the $D_s$ of observation on MJD 51094 is smaller than that on MJD 51032.
The event rate of J0942$-$5552 exhibited extreme fluctuations, spanning more than two orders of magnitude across the five epochs. The maximum event count (1,959 events) was recorded at MJD 51032, corresponding to an event rate of 0.620~$\rm events~period^{-1}$. Remarkably, after 64-day interval (MJD 51096), the event count decreased to just 14, with a rate of 0.004~$\rm events~period^{-1}$, representing only 0.6\% of the peak value~\footnote{Using the mean width of the 334 events in MJD 50986~(8.3ms), we obtain a fluence threshold of 0.86 Jy ms in our detection pipeline. If the 334 events are placed at a $D_s$ of 0.195, taking the gain degrading factor 0.439 into account, we found that 80.6\% pulses are still above the threshold. The 19.3\% of event rate losses is relatively small comparing to the observed times to orders event rate changes.}.

\begin{table}[H]
\begin{center}

\caption{The emission characteristic of different epochs of J0942-5552. The fluence is shown in $\rm Mean \pm Sigma$.\label{table:rateJ0942}}
\renewcommand\arraystretch{1.0}    
\begin{tabular}{ccccc}
\hline
MJD  &     Fluence        &     $\rm N_{events}$  	 &     	$ \rm R_{events}$   & $D_s$ 
\\
  &    (Jy ms)   &   &  $ \rm (events~period^{-1})$   &  (deg)  
\\\hline
50986 & $4.2 \pm 2.4$ & 334 & 0.583  & 0.000\\\hline
51032 & $4.8 \pm 2.4$ & 1959 & 0.620  & 0.131\\\hline
51094 & $2.6 \pm 0.7$ & 560 & 0.177  & 0.121\\\hline
51096 & $3.2 \pm 0.7$ & 14 & 0.004  & 0.194\\\hline
51100 & $3.1 \pm 1.0$ & 805 & 0.255  & 0.131\\\hline
\end{tabular}

\end{center}
\end{table}

\subsubsection{PSR J1453$-$6413}

\begin{figure*}
    \begin{center}
    \includegraphics[width=0.9\textwidth]{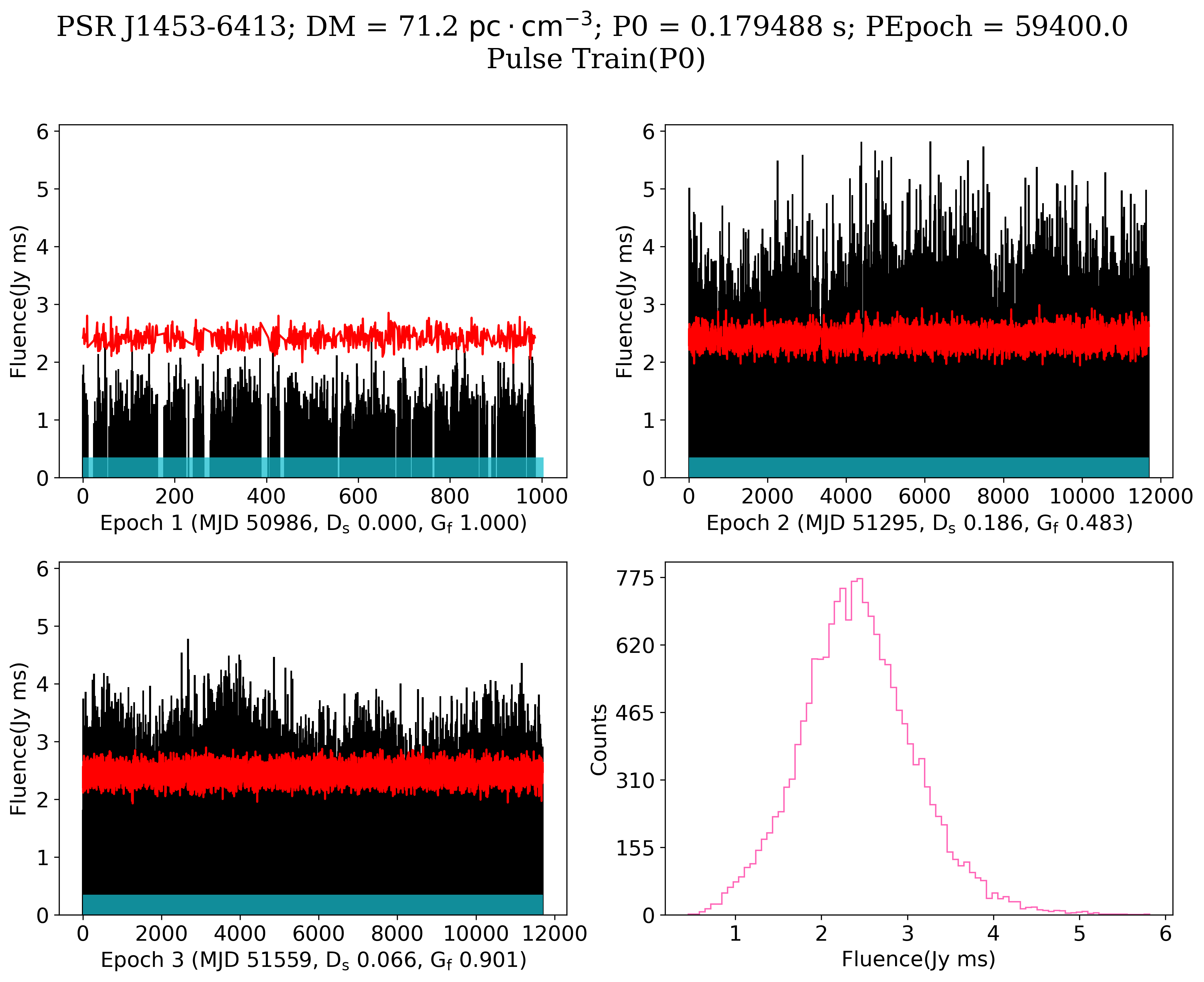}
    \caption{Fluence of 17,449 single pulses of PSR J1453$-$6413 from 4 observation epochs. The X-axis represents the pulse train number relative to the start of the observation, measured in units of the rotating period (P0). Each observation epoch is indicated by a blue background. The red solid lines represent the RMS level variation scaling to mean fluence level.}
    \label{figure:J1453}
    \end{center}
\end{figure*}

The emission characteristic of PSR J1453$-$6413 is shown in Table~\ref{table:rateJ1453} and figure~\ref{figure:J1453}. A comparative analysis of the data in Table~\ref{table:rateJ1453} reveals distinct flux evolution in J1453-6413 across the observed epochs. During the last two epochs (MJD 51559 and 51295), the fluence remained relatively stable at $2.7\pm0.7$~Jy ms and $2.3\pm0.6$~Jy ms. In contrast, the first epoch (MJD 50986) exhibited a marked decrease in fluence, $1.3\pm0.3$~Jy ms.
\begin{table}[H]
\begin{center}

\caption{The emission characteristic of different epochs of J1453-6413. The fluence is shown in $\rm Mean \pm Sigma$.\label{table:rateJ1453}}
\renewcommand\arraystretch{1.0}    
\begin{tabular}{ccccc}
\hline
MJD  &     Fluence        &     $\rm N_{events}$  	 &     	$ \rm R_{events}$   & $D_s$ 
\\
  &    (Jy ms)   &   &  $ \rm (events~period^{-1})$   &  (deg)  
\\\hline
50986 & $1.3 \pm 0.3$ & 740 & 0.737  & 0.000\\\hline
51295 & $2.7 \pm 0.7$ & 5846 & 0.500  & 0.186\\\hline
51559 & $2.3 \pm 0.6$ & 10863 & 0.928  & 0.066\\\hline
\end{tabular}

\end{center}
\end{table}

\subsection{Database applications}

\subsubsection{Advantages of the new database}

Unlike previous databases storing processed parameters~\citep{songbodatabase}, 
PTD II preserves raw data segments for each detected pulse. 
This allows direct analysis of original data without accessing full archives, requiring substantial storage and processing resources.
The database architecture reduces computational overhead by providing pre-identified pulse events. 
Traditional approaches process entire observation files (typically $\sim$100 MB each for Parkes multibeam observation), while PTD II reduces this to accessing KB-sized segments containing only the relevant pulses.
The SQLite3 implementation and standardised data structures ensure platform independence and integration potential with existing analysis tools, making the database accessible to researchers across different computing environments. Researchers can effectively manage their new data reduction pipelines through the PTD II. Special pulsars requiring raw data for emission behavior analysis benefit from the systematically organized storage in PTD II.

With 165,592 single pulses from 363 pulsars, PTD II provides sufficient statistical power to identify rare emission phenomena and establish robust emission pattern classifications across a significant pulsar population. 
This scale enables meaningful population-level studies that were impractical with smaller datasets. By collating pulses from observations spanning multiple epochs (1997-2001), the database facilitates investigations into long-term emission variability and mode-changing behaviour, as demonstrated in our examples of J1602-5100, J1539-5626, J0942-5552 and J1453-6413. This temporal dimension is often lacking in single-epoch studies. 
 
The database structure facilitates detailed analysis of pulse-to-pulse correlations and patterns that may reveal magnetospheric dynamics previously undetectable in integrated profiles. Such variations could provide constraints on models of plasma processes in extreme magnetic fields. The database can help identify and characterise previously undocumented transient emission phenomena. The ability to directly access raw data segments for these states enables detailed spectral and temporal analysis of these transitions. 

The standardised format of the database makes it an ideal training set for machine learning algorithms designed to detect and classify pulsar emission patterns. This could lead to the automated discovery of new emission classes or subtle pattern variations undetectable through traditional analysis methods.


\subsubsection{Future development of the database}

The growth of radio observational data volume presents unprecedented challenges as radio telescopes evolve. Two decades ago, the multibeam receiver system on the Parkes telescope generated approximately 500 MB of data per hour during search-mode observations. The FAST telescope has developed significantly, a single central beam of its 19-beam receiver produces 1 TB of data from 2-hour search-mode observations, with full 19-beam deployment multiplying this volume nineteen times~\citep{fast_19beam}. The upcoming Cryogenic Phased Array Feed~(CryoPAF) receiver, scheduled for commissioning in 2025 at the Parkes telescope, has 72 beams and a bandwidth of 700 MHz to 1950 MHz. Its search-mode surveys' data rates will reach TB per hour~\footnote{As cryoPAF is still being tested, the accurate parameters of the full-beam system performance are unavailable for now. We used the values referenced from the Parkes cryoPAF introduction by Alex Dunning: \url{https://research.csiro.au/ratechnologies/wp-content/uploads/sites/295/2022/11/PAFAR2022-Dunning-CryoPAF_for_Parkes.pdf}}. The Square Kilometre Array~(SKA) anticipates PB daily data outputs according to its white paper~\citep{ska1}. Under this big data challenge, the classical data storage mode will face difficulty in processing and retrieval efficiency. Our research focuses on integrating databases into radio data management systems. This framework aims to establish a unified database optimised for single-pulse studies (e.g., pulsars, FRBs), enabling more efficient scientific research of next-generation radio astronomy datasets.

For the ongoing advancement of our database, we will implement an upgrade plan comprising four key components. First, we will integrate the latest observation data from telescopes including the FAST and Parkes, incorporating both raw observational data and processed analysis results.
Second, we will use multiple searching pipeline like \emph{\sc PRESTO}, \emph{\sc HEIMDALL}\footnote{\url{https://sourceforge.net/projects/heimdall-astro/}} to cross-validate the single pulses. Third, our AI integration will expand to include neural networks for automated pulse classification. This will incorporate both supervised learning models trained on historical datasets~\citep{Yang21} and unsupervised anomaly detection systems~\citep{Yang25}. Finally, we will design a user interface with a responsive web framework supporting collaborative tools, data visualisation modules, and API endpoints for scripts access. 


\section*{DATA AVAILABILITY STATEMENTS}
The data used in this work is available from \url{https://data.csiro.au/}. PTD II, 363 pulsars' fluence supplementary material, and the python script is available from \url{https://github.com/Astroyx/Pulsar_collection}.

\section*{Acknowledgments}
This work is partially supported by the international Partnership Program of Chinese Academy of Sciences for Grand Challenges (114332KYSB20210018), the National SKA Program of China (2022SKA0130100), the National Natural Science Foundation of China (grant Nos. 12041306, 12273113, 12233002, 12003028), the National Key R\&D Program of China (2021YFA0718500), Postdoctoral Fellowship Program of CPSF (grant No. GZC20252100), the ACAMAR Postdoctoral Fellow, China Postdoctoral Science Foundation (grant No. 2025M773201), and Jiangsu Funding Program for Excellent Postdoctoral Talent. \\


\bibliography{main}
\bibliographystyle{aasjournal}

\section*{APPENDIX}

\subsection*{Downloading and using the database}

The database is available from~\footnote{As a large file (Pulsar\_fits\_database\_v1.zip) in this repository is managed by Git LFS (Large File Storage), cloning the repository or downloading it as a ZIP will only retrieve a small pointer file (usually a few kilobytes) instead of the actual large file. To access the full file, please install Git LFS and use git lfs pull after cloning the repository. Alternatively, you can click on the file in the GitHub web interface and use the "Download raw file" option to download the file directly.}:
\begin{equation*}
     \text{\url{https://github.com/Astroyx/Pulsar_collection.git}}
\end{equation*}
or
\begin{equation*}
     \text{\url{https://doi.org/10.25919/34am-zx04}.}
\end{equation*}

Currently:
\begin{itemize}
\item it is distributed as a zip file (Pulsar\_database.zip),the archive is $\sim$ 1.5 GB;

\item when extracted the SQLite database using \emph{\sc unzip}, requires $\sim$ 4 GB for the single database file~(Pulsar\_fits\_database.db);

\item The database can be accessed using \emph{\sc SQLite} directly; 

\item We also provide a software tool to manage the database:\emph{\sc get\_pulsar\_pub.py}, available from: \begin{equation*}
     \text{\url{https://github.com/Astroyx/Pulsar_collection};}
\end{equation*}

\item \emph{\sc get\_pulsar\_pub.py} can extract the information and all file segments of a pulsar with Jname input;
\begin{itemize}
\item use the
\begin{equation*}
    \text{python get\_pulsar\_pub.py -h} 
\end{equation*}
command to print the help page;
\item use the
\begin{equation*}
    \text{ python get\_pulsar\_pub.py -db Pulsar\_fits\_database.db -jlist }
\end{equation*}
command to print all the pulsar name in the database;
\item use the
\begin{equation*}
     \text{python get\_pulsar\_pub.py -db  Pulsar\_fits\_database.db -fluence -j J1745-3040 }
\end{equation*}
command to extract all the file segments and fluence fitting result of PSR J1745$-$3040 in the database.
\item use the
\begin{equation*}
     \text{python get\_pulsar\_pub.py -db  Pulsar\_fits\_database.db -fluence -j J1745-3040 -mjd 50690}
\end{equation*}
command to extract all the file segments and fluence fitting result when the observation MJD of PSR J1745$-$3040 is 50690 in the database.
\end{itemize}
\end{itemize}
\clearpage
\subsection*{The 363 pulsars fluence distribution figure set}
\begin{figure*}[htbp]
    \begin{center}
    \includegraphics[width=0.9\textwidth]{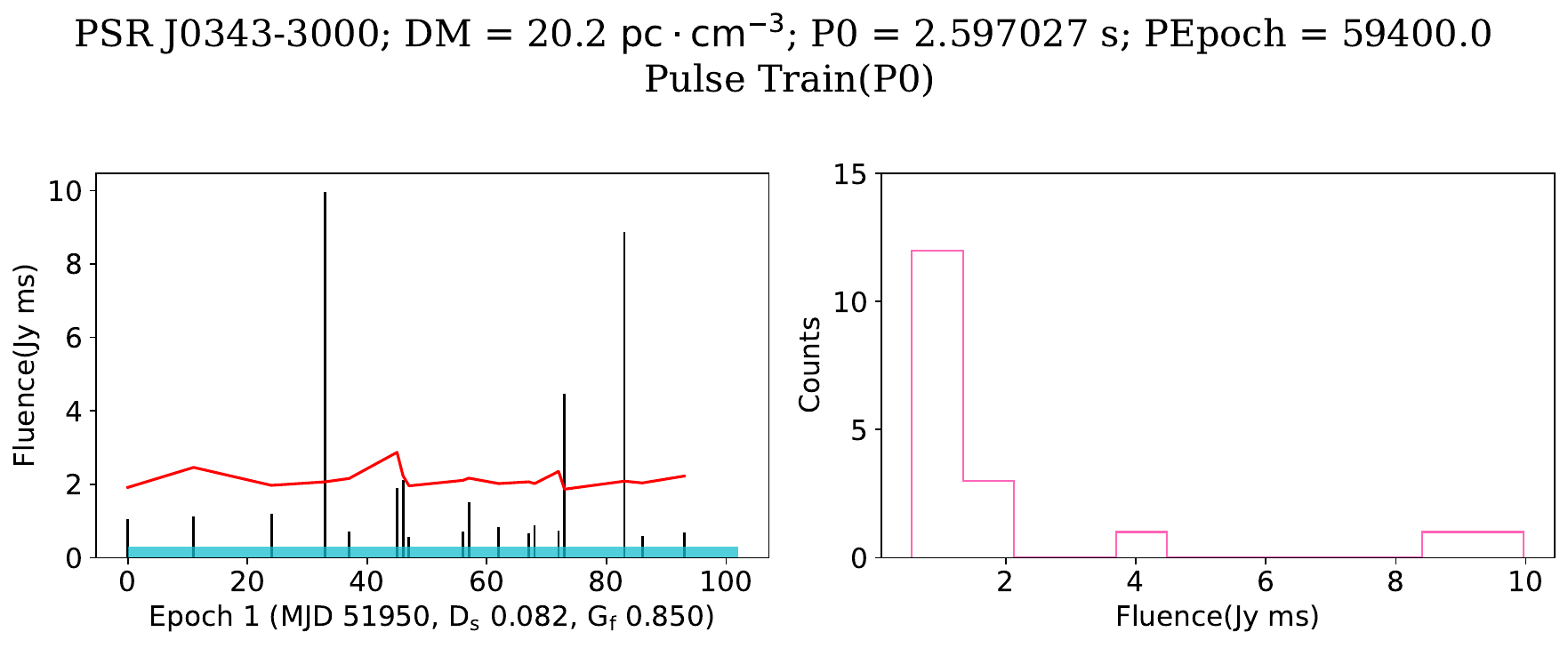}
    \caption{Fluence of 18 single pulses of PSR J0343$-$3000 from 1 observation epoch. The X-axis represents the pulse train number relative to the start of the observation, measured in units of the rotating period (P0). Each observation epoch is indicated by a blue background. The red solid lines represent the RMS level variation scaling to mean fluence level. The complete figure set (363 images) is available in the online journal.}
    \label{figure:J0343}
    \end{center}
\end{figure*}

\clearpage

\subsection*{Single-pulse Table of 363 pulsars}

\begin{table*}[ht]
  \begin{scriptsize}
  \caption{Single-pulse detections of Pulsars. Columns are (1) PSR name (J2000); (2) Period of the pulsar; (3) DM of the pulsar; (4) number of pulses detected in the single-pulse searching pipeline, and total number of periods in the observation; (5) S/N of the brightest single pulse detected in the single-pulse searching pipeline; (6) S/N of the faintest single pulse detected in the single-pulse searching pipeline under a detection threshold of 8; (7) The fluence ratio is defined as the pulse exhibiting the largest fluence divided by the pulse with the smallest fluence; (8) We divided the pulsars into three energy distribution classifications if more than 100 single pulses are detected: Log-normal~(L), Gaussian~(G), Unimodal~(U) and Other~(O). \label{table:properties}}
  \renewcommand\arraystretch{1.0}
  \setlength{\tabcolsep}{2.0mm}  
  \begin{center}
  \begin{tabular}{@{\extracolsep{\fill}}cccccccc}
  \hline
\multicolumn{1}{c}{Jname}  &    Period      &  DM        &     Number of Single     &     Max SP         &    Min SP  & Fluence  & Distribution         \\
 \multicolumn{1}{c}{}    &   (s)     &   (pc$\cdot$cm$^{-3}$)         &    Pulses Detected         &        S/N       &  S/N    & ratio &   class     \\
\hline
J0343$-$3000 & 2.5970 & 20.20 & 19/204 & 13.41 & 8.27 & 17.8 & -\\ 
J0410$-$31 & 1.8785 & 9.20 & 1/282 & 12.16 & 12.16 & 1.0 & -\\ 
J0418$-$4154 & 0.7571 & 24.32 & 2/1401 & 11.14 & 8.76 & 1.7 & -\\ 
J0529$-$6652 & 0.9757 & 103.31 & 6/33158 & 9.60 & 8.05 & 3.4 & -\\ 
J0540$-$6919 & 0.0506 & 147.20 & 2/849715 & 10.11 & 8.02 & 1.0 & -\\ 
J0726$-$2612 & 3.4423 & 69.40 & 8/308 & 18.82 & 8.61 & 6.4 & -\\ 
J0729$-$1836 & 0.5102 & 61.29 & 699/9792 & 22.80 & 8.01 & 14.7 & O\\ 
J0736$-$6304 & 4.8629 & 19.40 & 2/163 & 8.57 & 8.23 & 1.0 & -\\ 
J0738$-$4042 & 0.3749 & 160.70 & 625/707 & 25.19 & 8.04 & 11.9 & O\\ 
J0742$-$2822 & 0.1668 & 73.76 & 71/68960 & 10.22 & 8.02 & 2.9 & -\\ 
J0745$-$5353 & 0.2148 & 121.38 & 12/3703 & 12.37 & 8.04 & 2.2 & -\\ 
J0758$-$1528 & 0.6823 & 63.23 & 153/1166 & 16.94 & 8.01 & 5.5 & -\\ 
J0809$-$4753 & 0.5472 & 228.68 & 1/2908 & 8.17 & 8.17 & 1.0 & -\\ 
J0818$-$3232 & 2.1613 & 131.80 & 21/245 & 11.69 & 8.10 & 3.8 & -\\ 
J0820$-$1350 & 1.2381 & 40.94 & 37/428 & 18.27 & 8.02 & 13.2 & -\\ 
J0820$-$4114 & 0.5454 & 113.40 & 1/8673 & 8.11 & 8.11 & 1.0 & -\\ 
J0823+0159 & 0.8649 & 23.73 & 52/919 & 14.81 & 8.02 & 10.0 & -\\ 
J0828$-$3417 & 1.8489 & 51.60 & 8/1709 & 13.61 & 9.10 & 7.6 & -\\ 
J0835$-$3707 & 0.5414 & 112.40 & 10/5838 & 15.69 & 8.33 & 1.8 & -\\ 
J0837$-$4135 & 0.7516 & 147.22 & 16273/19559 & 44.40 & 8.01 & 78.8 & O\\ 
J0842$-$4851 & 0.6444 & 196.69 & 56/13037 & 15.40 & 8.06 & 5.8 & -\\ 
J0846$-$3533 & 1.1161 & 94.16 & 65/4714 & 13.27 & 8.02 & 10.5 & -\\ 
J0904$-$4246 & 0.9652 & 145.80 & 74/6528 & 19.07 & 8.06 & 6.1 & -\\ 
J0905$-$4536 & 0.9883 & 196.22 & 2/12750 & 10.54 & 8.65 & 1.2 & -\\ 
J0907$-$5157 & 0.2536 & 103.64 & 5820/41415 & 25.23 & 8.01 & 18.3 & O\\ 
J0908$-$4913 & 0.1068 & 180.13 & 625/118022 & 11.13 & 8.01 & 19.7 & -\\ 
J0922+0638 & 0.4306 & 27.23 & 321/615 & 21.36 & 8.01 & 19.3 & L\\ 
J0922$-$4949 & 0.9504 & 237.30 & 4/15469 & 9.43 & 8.06 & 1.2 & -\\ 
J0924$-$5302 & 0.7463 & 153.38 & 2/8442 & 8.58 & 8.55 & 2.5 & -\\ 
J0924$-$5814 & 0.7395 & 57.10 & 144/8520 & 13.08 & 8.01 & 7.4 & L\\ 
J0934$-$5249 & 1.4448 & 100.34 & 108/5814 & 16.68 & 8.03 & 5.1 & O\\ 
J0941$-$39 & 0.5868 & 78.00 & 7/1807 & 21.47 & 8.16 & 3.2 & -\\ 
J0942$-$5552 & 0.6644 & 180.20 & 3675/19416 & 32.37 & 8.01 & 32.7 & O\\ 
J0953+0755 & 0.2531 & 2.94 & 435/4192 & 34.36 & 8.01 & 178.9 & O\\ 
J0955$-$5304 & 0.8621 & 156.73 & 8/9744 & 11.38 & 8.05 & 2.6 & -\\ 
J1001$-$5507 & 1.4366 & 130.27 & 4453/5951 & 35.54 & 8.01 & 34.9 & O\\ 
J1001$-$5939 & 7.7336 & 113.00 & 89/1086 & 19.99 & 8.08 & 6.1 & -\\ 
J1012$-$5857 & 0.8199 & 382.61 & 85/5305 & 26.10 & 8.01 & 3.6 & -\\ 
J1020$-$5921 & 1.2383 & 80.00 & 8/3392 & 11.95 & 8.04 & 5.3 & -\\ 
J1022+1001 & 0.0165 & 10.26 & 104/32239 & 11.20 & 8.01 & 4.7 & -\\ 
J1032$-$5206 & 2.4076 & 139.00 & 1/1202 & 8.50 & 8.50 & 1.0 & -\\ 
J1032$-$5911 & 0.4642 & 418.20 & 67/22621 & 10.02 & 8.03 & 3.5 & -\\ 
J1038$-$5831 & 0.6620 & 72.78 & 172/16316 & 16.42 & 8.01 & 11.5 & -\\ 
J1042$-$5521 & 1.1709 & 306.74 & 30/8968 & 10.47 & 8.12 & 3.0 & -\\ 
J1043$-$6116 & 0.2886 & 448.75 & 242/21831 & 16.57 & 8.01 & 21.5 & O\\ 
J1047$-$6709 & 0.1985 & 116.30 & 142/7032 & 25.98 & 8.06 & 8.7 & -\\ 
J1048$-$5832 & 0.1237 & 128.68 & 7726/87280 & 23.65 & 8.01 & 43.1 & O\\ 
J1048$-$5838 & 1.2313 & 70.70 & 14/7065 & 23.61 & 8.17 & 10.3 & -\\ 
J1049$-$5833 & 2.2023 & 448.81 & 62/4903 & 13.13 & 8.01 & 5.4 & -\\ 
J1055$-$6905 & 2.9194 & 142.80 & 59/4476 & 12.34 & 8.03 & 5.2 & -\\ 
J1056$-$6258 & 0.4225 & 320.69 & 573/19886 & 16.43 & 8.01 & 19.6 & -\\ 
J1057$-$5226 & 0.1971 & 29.69 & 444/7387 & 15.61 & 8.01 & 22.0 & O\\ 
J1059$-$5742 & 1.1850 & 108.56 & 547/14178 & 19.83 & 8.01 & 9.9 & O\\ 
J1104$-$6103 & 0.2809 & 78.51 & 3/25849 & 9.08 & 8.20 & 1.7 & -\\ 
J1106$-$6438 & 2.7179 & 203.00 & 1/3090 & 8.19 & 8.19 & 1.0 & -\\ 
J1107$-$5907 & 0.2528 & 40.75 & 7/49852 & 14.83 & 8.27 & 3.4 & -\\ 
J1110$-$5637 & 0.5583 & 261.68 & 65/23110 & 14.12 & 8.01 & 5.0 & -\\ 
J1112$-$6926 & 0.8205 & 148.67 & 1/969 & 8.09 & 8.09 & 1.0 & -\\ 
J1114$-$6100 & 0.8809 & 680.06 & 92/10739 & 22.49 & 8.07 & 7.8 & -\\ 
J1116$-$4122 & 0.9432 & 40.53 & 72/843 & 16.95 & 8.06 & 14.3 & -\\ 
\hline
\end{tabular}
\end{center}
\end{scriptsize}
\end{table*}
\begin{table*}[ht]
    \renewcommand\thetable{\ref{table:properties}}
  \begin{scriptsize}
  \caption{Continued.}
  \renewcommand\arraystretch{1.0}
  \setlength{\tabcolsep}{2.0mm}  
  \begin{center}
  \begin{tabular}{@{\extracolsep{\fill}}cccccccc}
  \hline  
\multicolumn{1}{c}{Jname}  &    Period      &  DM        &     Number of Single     &     Max SP         &    Min SP  & Fluence  & Distribution         \\
 \multicolumn{1}{c}{}    &   (s)     &   (pc$\cdot$cm$^{-3}$)         &    Pulses Detected         &        S/N       &  S/N    & ratio &   class     \\
\hline
J1117$-$6154 & 0.5051 & 493.38 & 1/14255 & 8.26 & 8.26 & 1.0 & -\\ 
J1119$-$7936 & 2.2806 & 26.32 & 1/465 & 9.99 & 9.99 & 1.0 & -\\ 
J1129$-$53 & 1.0629 & 77.00 & 2/998 & 16.73 & 9.47 & 4.5 & -\\ 
J1133$-$6250 & 1.0229 & 568.46 & 116/10442 & 17.40 & 8.02 & 4.0 & -\\ 
J1136+1551 & 1.1879 & 4.78 & 22/31 & 12.11 & 8.03 & 26.6 & -\\ 
J1136$-$5525 & 0.3647 & 85.11 & 44/2181 & 13.14 & 8.01 & 4.2 & -\\ 
J1141$-$6545 & 0.3939 & 116.08 & 2833/16910 & 25.31 & 8.01 & 8.7 & O\\ 
J1144$-$6146 & 0.9878 & 78.70 & 1/11238 & 8.05 & 8.05 & 1.0 & -\\ 
J1146$-$6030 & 0.2734 & 111.68 & 612/30730 & 22.81 & 8.01 & 7.5 & L\\ 
J1148$-$5725 & 3.5599 & 174.00 & 11/1769 & 20.77 & 8.01 & 8.6 & -\\ 
J1157$-$6224 & 0.4005 & 324.40 & 2504/26218 & 28.48 & 8.01 & 25.8 & O\\ 
J1159$-$7910 & 0.5251 & 59.32 & 4/1515 & 13.63 & 8.94 & 2.4 & -\\ 
J1202$-$5820 & 0.4528 & 145.74 & 38/13914 & 11.84 & 8.02 & 2.7 & -\\ 
J1224$-$6407 & 0.2165 & 97.73 & 2816/29658 & 25.07 & 8.01 & 7.3 & O\\ 
J1239$-$6832 & 1.3019 & 94.12 & 223/5246 & 26.33 & 8.02 & 11.7 & O\\ 
J1243$-$6423 & 0.3885 & 297.02 & 2736/21776 & 34.92 & 8.01 & 29.5 & O\\ 
J1252$-$6314 & 0.8233 & 276.85 & 4/10203 & 10.64 & 9.07 & 1.5 & -\\ 
J1253$-$5820 & 0.2555 & 100.55 & 49/16440 & 11.30 & 8.06 & 6.3 & -\\ 
J1255$-$6131 & 0.6580 & 206.50 & 8/9575 & 10.59 & 8.24 & 2.0 & -\\ 
J1306$-$6617 & 0.4730 & 438.23 & 12/17759 & 9.84 & 8.07 & 1.6 & -\\ 
J1307$-$6318 & 4.9624 & 374.00 & 8/1692 & 11.63 & 8.11 & 2.8 & -\\ 
J1307$-$67 & 3.6512 & 44.00 & 1/2300 & 8.81 & 8.81 & 1.0 & -\\ 
J1312$-$5516 & 0.8492 & 134.10 & 46/1249 & 13.44 & 8.01 & 5.8 & -\\ 
J1314$-$6101 & 2.9484 & 309.00 & 38/3561 & 15.00 & 8.01 & 3.1 & -\\ 
J1317$-$5759 & 2.6422 & 145.30 & 3/794 & 19.04 & 9.53 & 4.1 & -\\ 
J1319$-$6105 & 0.4211 & 440.60 & 13/24936 & 11.02 & 8.03 & 5.3 & -\\ 
J1320$-$3512 & 0.4585 & 16.81 & 15/393 & 15.35 & 8.06 & 3.9 & -\\ 
J1320$-$5359 & 0.2797 & 97.32 & 35/5080 & 12.40 & 8.02 & 2.8 & -\\ 
J1324$-$6302 & 2.4838 & 497.00 & 2/5073 & 11.92 & 10.27 & 4.1 & -\\ 
J1326$-$5859 & 0.4780 & 287.13 & 884/13432 & 13.68 & 8.01 & 18.7 & L\\ 
J1326$-$6408 & 0.7927 & 502.28 & 71/10598 & 10.90 & 8.01 & 9.2 & -\\ 
J1326$-$6700 & 0.5430 & 208.92 & 3329/15470 & 29.79 & 8.01 & 20.6 & O\\ 
J1327$-$6222 & 0.5299 & 318.52 & 8778/23779 & 27.08 & 8.01 & 58.3 & O\\ 
J1327$-$6301 & 0.1965 & 294.24 & 85/64135 & 13.01 & 8.05 & 3.4 & -\\ 
J1327$-$6400 & 0.2807 & 679.00 & 29/44896 & 11.96 & 8.05 & 4.6 & -\\ 
J1328$-$4357 & 0.5327 & 41.31 & 29/2668 & 10.91 & 8.02 & 4.8 & -\\ 
J1338$-$6204 & 1.2390 & 642.28 & 6/6780 & 11.22 & 8.21 & 2.0 & -\\ 
J1340$-$6456 & 0.3786 & 76.83 & 34/22187 & 16.94 & 8.03 & 4.5 & -\\ 
J1341$-$6023 & 0.6273 & 364.95 & 4/10044 & 11.38 & 8.94 & 1.4 & -\\ 
J1345$-$6115 & 1.2531 & 280.47 & 1/8380 & 10.65 & 10.65 & 1.0 & -\\ 
J1347$-$5947 & 0.6100 & 292.34 & 4/7208 & 9.74 & 8.36 & 2.0 & -\\ 
J1349$-$6130 & 0.2594 & 283.87 & 1/24798 & 8.94 & 8.94 & 1.0 & -\\ 
J1355$-$5747 & 2.0387 & 229.00 & 1/3090 & 8.14 & 8.14 & 1.0 & -\\ 
J1357$-$62 & 0.4558 & 416.70 & 2/18432 & 10.35 & 8.05 & 1.3 & -\\ 
J1357$-$6429 & 0.1663 & 128.50 & 1/37885 & 10.04 & 10.04 & 1.0 & -\\ 
J1359$-$6038 & 0.1275 & 293.91 & 2557/100224 & 13.77 & 8.01 & 18.8 & O\\ 
J1401$-$6357 & 0.8428 & 97.74 & 1744/9967 & 27.70 & 8.01 & 13.9 & O\\ 
J1405$-$5641 & 0.6176 & 273.00 & 2/11061 & 9.00 & 8.39 & 1.5 & -\\ 
J1406$-$5806 & 0.2883 & 229.86 & 662/29134 & 26.94 & 8.01 & 15.7 & L\\ 
J1413$-$6307 & 0.3950 & 121.63 & 58/15952 & 18.88 & 8.03 & 2.5 & -\\ 
J1414$-$6802 & 4.6302 & 153.50 & 5/454 & 10.73 & 8.13 & 1.3 & -\\ 
J1416$-$6037 & 0.2956 & 288.88 & 19/14210 & 13.45 & 8.12 & 3.1 & -\\ 
J1423$-$6953 & 0.3334 & 123.98 & 39/3181 & 21.13 & 8.01 & 3.8 & -\\ 
J1424$-$56 & 1.4270 & 32.90 & 5/5887 & 21.70 & 10.12 & 3.0 & -\\ 
J1428$-$5530 & 0.5703 & 81.92 & 2971/14730 & 27.60 & 8.01 & 22.4 & O\\ 
J1430$-$6623 & 0.7854 & 65.12 & 2348/6360 & 25.49 & 8.01 & 27.4 & O\\ 
J1444$-$6026 & 4.7586 & 367.70 & 2/2292 & 10.64 & 9.43 & 1.7 & -\\ 
J1452$-$6036 & 0.1550 & 350.00 & 199/54202 & 16.26 & 8.02 & 12.1 & O\\ 
J1453$-$6413 & 0.1795 & 71.18 & 17451/36107 & 27.05 & 8.01 & 12.9 & O\\ 
J1456$-$6843 & 0.2634 & 8.63 & 501/4482 & 13.31 & 8.01 & 85.9 & L\\ 
\hline
\end{tabular}
\end{center}
\end{scriptsize}
\end{table*}
\begin{table*}[ht]
    \renewcommand\thetable{\ref{table:properties}}
  \begin{scriptsize}
  \caption{Continued.}
  \renewcommand\arraystretch{1.0}
  \setlength{\tabcolsep}{2.0mm}  
  \begin{center}
  \begin{tabular}{@{\extracolsep{\fill}}cccccccc}
  \hline  
\multicolumn{1}{c}{Jname}  &    Period      &  DM        &     Number of Single     &     Max SP         &    Min SP  & Fluence  & Distribution         \\
 \multicolumn{1}{c}{}    &   (s)     &   (pc$\cdot$cm$^{-3}$)         &    Pulses Detected         &        S/N       &  S/N    & ratio &   class     \\
\hline
J1457$-$5122 & 1.7483 & 37.00 & 27/606 & 16.98 & 8.12 & 10.1 & -\\ 
J1502$-$5653 & 0.5355 & 193.40 & 149/19609 & 15.99 & 8.01 & 2.9 & -\\ 
J1504$-$5621 & 0.4130 & 148.60 & 1/15256 & 8.99 & 8.99 & 1.0 & -\\ 
J1507$-$4352 & 0.2868 & 49.13 & 267/9248 & 15.04 & 8.01 & 4.2 & -\\ 
J1507$-$6640 & 0.3557 & 130.21 & 18/2982 & 12.86 & 8.11 & 2.2 & -\\ 
J1512$-$5759 & 0.1287 & 627.47 & 1/32637 & 8.50 & 8.50 & 1.0 & -\\ 
J1513$-$5946 & 1.0461 & 171.70 & 7/6022 & 13.43 & 8.02 & 2.8 & -\\ 
J1519$-$6106 & 2.1543 & 221.00 & 5/974 & 9.82 & 8.20 & 1.7 & -\\ 
J1519$-$6308 & 1.2541 & 249.82 & 1/6987 & 8.12 & 8.12 & 1.0 & -\\ 
J1522$-$5525 & 1.3896 & 79.00 & 1/4534 & 9.24 & 9.24 & 1.0 & -\\ 
J1522$-$5829 & 0.3954 & 199.69 & 1/16695 & 8.04 & 8.04 & 1.0 & -\\ 
J1527$-$3931 & 2.4176 & 48.80 & 2/548 & 9.20 & 8.94 & 1.5 & -\\ 
J1528$-$4109 & 0.5266 & 89.28 & 9/14913 & 10.43 & 8.03 & 1.9 & -\\ 
J1534$-$46 & 0.3648 & 64.00 & 3/6542 & 8.69 & 8.28 & 1.2 & -\\ 
J1534$-$5334 & 1.3689 & 25.26 & 1953/6137 & 21.11 & 8.01 & 20.2 & O\\ 
J1535$-$4114 & 0.4329 & 66.28 & 758/8636 & 20.95 & 8.01 & 21.0 & L\\ 
J1536$-$3602 & 1.3198 & 86.45 & 10/803 & 20.04 & 8.03 & 2.1 & -\\ 
J1536$-$5433 & 0.8814 & 147.50 & 65/11913 & 14.56 & 8.01 & 6.3 & -\\ 
J1537$-$4912 & 0.3013 & 69.70 & 91/39135 & 17.47 & 8.01 & 16.0 & -\\ 
J1539$-$5626 & 0.2434 & 175.97 & 214/26378 & 16.82 & 8.01 & 4.9 & L\\ 
J1542$-$5034 & 0.5992 & 89.54 & 4/7452 & 9.64 & 8.61 & 1.7 & -\\ 
J1542$-$5303 & 1.2076 & 265.70 & 14/5217 & 14.70 & 8.34 & 3.9 & -\\ 
J1548$-$4927 & 0.6027 & 140.75 & 83/10453 & 16.44 & 8.01 & 3.7 & -\\ 
J1549$-$4848 & 0.2884 & 56.21 & 3/36935 & 12.08 & 10.94 & 1.1 & -\\ 
J1553$-$5456 & 1.0813 & 210.00 & 1/9711 & 8.72 & 8.72 & 1.0 & -\\ 
J1557$-$4258 & 0.3292 & 144.38 & 106/2459 & 16.34 & 8.01 & 5.3 & -\\ 
J1558$-$5756 & 1.1223 & 127.80 & 20/7485 & 12.45 & 8.26 & 1.9 & -\\ 
J1559$-$4438 & 0.2571 & 56.09 & 3246/3561 & 28.18 & 8.01 & 27.1 & G\\ 
J1559$-$5545 & 0.9573 & 212.90 & 15/11283 & 11.35 & 8.04 & 2.9 & -\\ 
J1600$-$5044 & 0.1926 & 262.87 & 2/54521 & 8.47 & 8.20 & 1.3 & -\\ 
J1602$-$5100 & 0.8643 & 170.79 & 2068/7428 & 28.92 & 8.01 & 14.6 & O\\ 
J1603$-$2531 & 0.2831 & 53.76 & 456/2445 & 26.02 & 8.01 & 12.1 & L\\ 
J1603$-$2712 & 0.7783 & 46.00 & 6/917 & 9.35 & 8.23 & 2.9 & -\\ 
J1603$-$3539 & 0.1419 & 78.39 & 2/14951 & 8.53 & 8.42 & 1.2 & -\\ 
J1604$-$4909 & 0.3274 & 140.75 & 1528/19243 & 14.91 & 8.01 & 15.4 & L\\ 
J1605$-$5257 & 0.6580 & 34.80 & 1192/28725 & 17.73 & 8.01 & 71.7 & L\\ 
J1612$-$2408 & 0.9238 & 49.20 & 1/1311 & 8.87 & 8.87 & 1.0 & -\\ 
J1615$-$5444 & 0.3610 & 312.60 & 1/17455 & 8.75 & 8.75 & 1.0 & -\\ 
J1615$-$5537 & 0.7915 & 124.48 & 2/13266 & 8.98 & 8.72 & 1.1 & -\\ 
J1617$-$4216 & 3.4285 & 164.49 & 3/5576 & 9.76 & 9.58 & 1.1 & -\\ 
J1624$-$4613 & 0.8712 & 224.20 & 14/9642 & 16.03 & 8.03 & 3.8 & -\\ 
J1626$-$4537 & 0.3701 & 236.59 & 10/22696 & 10.34 & 8.04 & 1.8 & -\\ 
J1632$-$4621 & 1.7092 & 562.02 & 1/2457 & 8.63 & 8.63 & 1.0 & -\\ 
J1633$-$4453 & 0.4365 & 472.22 & 5/14434 & 8.71 & 8.01 & 2.0 & -\\ 
J1633$-$5015 & 0.3521 & 399.06 & 622/29820 & 13.07 & 8.01 & 31.8 & O\\ 
J1635$-$4513 & 1.5947 & 416.00 & 1/2633 & 8.25 & 8.25 & 1.0 & -\\ 
J1641$-$2347 & 1.0910 & 27.70 & 4/5880 & 9.75 & 8.15 & 2.2 & -\\ 
J1646$-$6831 & 1.7856 & 41.96 & 44/742 & 18.58 & 8.05 & 18.9 & -\\ 
J1647$-$3607 & 0.2123 & 228.50 & 5/3747 & 11.33 & 8.70 & 3.4 & -\\ 
J1649$-$4349 & 0.8707 & 396.38 & 2/9648 & 8.16 & 8.09 & 1.1 & -\\ 
J1651$-$4246 & 0.8441 & 483.30 & 583/7464 & 17.64 & 8.01 & 16.5 & L\\ 
J1651$-$5222 & 0.6351 & 178.27 & 440/11174 & 13.71 & 8.01 & 19.7 & L\\ 
J1651$-$5255 & 0.8905 & 164.00 & 1/297 & 8.42 & 8.42 & 1.0 & -\\ 
J1652$-$2404 & 1.7037 & 68.41 & 22/934 & 19.16 & 8.03 & 3.5 & -\\ 
J1653$-$3838 & 0.3050 & 206.54 & 494/20655 & 18.21 & 8.02 & 16.7 & L\\ 
J1653$-$4249 & 0.6126 & 416.10 & 6/13714 & 8.71 & 8.03 & 1.7 & -\\ 
J1654$-$23 & 0.5454 & 75.00 & 3/3890 & 8.96 & 8.01 & 1.8 & -\\ 
J1655$-$3048 & 0.5429 & 154.30 & 1/17940 & 8.88 & 8.88 & 1.0 & -\\ 
J1700$-$3312 & 1.3583 & 166.70 & 20/1936 & 10.87 & 8.03 & 4.0 & -\\ 
J1701$-$3130 & 0.2913 & 130.73 & 1/21109 & 8.28 & 8.28 & 1.0 & -\\ 
\hline
\end{tabular}
\end{center}
\end{scriptsize}
\end{table*}
\begin{table*}[ht]
    \renewcommand\thetable{\ref{table:properties}}
  \begin{scriptsize}
  \caption{Continued.}
  \renewcommand\arraystretch{1.0}
  \setlength{\tabcolsep}{2.0mm}  
  \begin{center}
  \begin{tabular}{@{\extracolsep{\fill}}cccccccc}
  \hline  
\multicolumn{1}{c}{Jname}  &    Period      &  DM        &     Number of Single     &     Max SP         &    Min SP  & Fluence  & Distribution         \\
 \multicolumn{1}{c}{}    &   (s)     &   (pc$\cdot$cm$^{-3}$)         &    Pulses Detected         &        S/N       &  S/N    & ratio &   class     \\
\hline
J1701$-$3726 & 2.4546 & 303.40 & 577/3422 & 23.07 & 8.01 & 36.3 & O\\ 
J1703$-$3241 & 1.2118 & 110.09 & 2191/5199 & 25.46 & 8.01 & 25.3 & O\\ 
J1703$-$4442 & 1.7473 & 280.20 & 3/3605 & 13.58 & 8.15 & 2.2 & -\\ 
J1703$-$4851 & 1.3964 & 149.63 & 60/4512 & 17.37 & 8.05 & 4.5 & -\\ 
J1705$-$3423 & 0.2554 & 146.15 & 78/24667 & 10.89 & 8.01 & 3.1 & -\\ 
J1705$-$3950 & 0.3190 & 206.86 & 21/19752 & 15.82 & 8.07 & 2.6 & -\\ 
J1706$-$6118 & 0.3619 & 76.13 & 71/3664 & 23.78 & 8.11 & 4.6 & -\\ 
J1707$-$4053 & 0.5810 & 351.78 & 80/14915 & 11.57 & 8.02 & 5.8 & -\\ 
J1707$-$4417 & 5.7638 & 380.00 & 2/1457 & 10.03 & 8.33 & 1.1 & -\\ 
J1707$-$4729 & 0.2665 & 268.30 & 15/31526 & 9.74 & 8.03 & 4.4 & -\\ 
J1708$-$3426 & 0.6921 & 188.70 & 64/12137 & 17.19 & 8.03 & 5.4 & -\\ 
J1708$-$4522 & 1.2978 & 454.00 & 1/4854 & 10.12 & 10.12 & 1.0 & -\\ 
J1709$-$1640 & 0.6531 & 24.89 & 721/1624 & 23.03 & 8.01 & 26.8 & G\\ 
J1709$-$4429 & 0.1025 & 75.58 & 189/105343 & 14.45 & 8.01 & 5.4 & -\\ 
J1714$-$1054 & 2.0888 & 51.50 & 34/507 & 19.79 & 8.09 & 6.4 & -\\ 
J1715$-$4034 & 2.0722 & 254.00 & 43/3040 & 16.05 & 8.04 & 3.5 & -\\ 
J1717$-$3425 & 0.6563 & 583.46 & 560/12800 & 12.17 & 8.01 & 28.6 & L\\ 
J1717$-$4043 & 0.3979 & 452.60 & 7/21115 & 9.90 & 8.03 & 1.6 & -\\ 
J1720$-$2933 & 0.6204 & 42.75 & 304/10155 & 14.81 & 8.01 & 7.5 & L\\ 
J1721$-$3532 & 0.2804 & 496.82 & 288/37447 & 14.24 & 8.01 & 21.0 & O\\ 
J1722$-$3207 & 0.4772 & 126.09 & 397/13204 & 22.62 & 8.01 & 5.3 & L\\ 
J1722$-$3632 & 0.3992 & 415.67 & 3/26306 & 8.90 & 8.10 & 1.5 & -\\ 
J1723$-$3659 & 0.2027 & 254.40 & 21/22497 & 11.74 & 8.02 & 1.8 & -\\ 
J1725$-$4043 & 1.4651 & 203.00 & 6/4300 & 10.59 & 8.10 & 1.9 & -\\ 
J1727$-$2739 & 1.2931 & 146.00 & 274/6496 & 26.80 & 8.01 & 29.7 & L\\ 
J1730$-$3350 & 0.1395 & 261.29 & 36/77424 & 14.84 & 8.01 & 3.7 & -\\ 
J1731$-$4744 & 0.8299 & 122.85 & 1365/1712 & 37.28 & 8.01 & 51.1 & O\\ 
J1733$-$3716 & 0.3376 & 153.32 & 436/18663 & 30.31 & 8.01 & 6.9 & O\\ 
J1733$-$5515 & 1.0112 & 83.90 & 1/1311 & 8.09 & 8.09 & 1.0 & -\\ 
J1735$-$0724 & 0.4193 & 73.38 & 2/2529 & 8.55 & 8.46 & 1.3 & -\\ 
J1736$-$2457 & 2.6422 & 169.40 & 3/2384 & 8.39 & 8.11 & 1.7 & -\\ 
J1736$-$2843 & 6.4450 & 331.00 & 10/1303 & 21.29 & 8.14 & 3.6 & -\\ 
J1737$-$3555 & 0.3976 & 89.18 & 11/26411 & 10.97 & 8.02 & 2.4 & -\\ 
J1738$-$2330 & 1.9788 & 96.60 & 6/3184 & 12.36 & 8.01 & 13.6 & -\\ 
J1738$-$3211 & 0.7685 & 49.34 & 93/8198 & 17.74 & 8.01 & 7.1 & -\\ 
J1739$-$1313 & 1.2157 & 58.20 & 131/5021 & 12.49 & 8.01 & 8.1 & -\\ 
J1739$-$2521 & 1.8185 & 186.40 & 2/4619 & 13.18 & 8.25 & 3.4 & -\\ 
J1739$-$2903 & 0.3229 & 138.66 & 240/27134 & 21.06 & 8.01 & 5.1 & -\\ 
J1739$-$3131 & 0.5294 & 596.90 & 52/19834 & 10.31 & 8.01 & 15.9 & -\\ 
J1740$-$3015 & 0.6071 & 151.85 & 3447/10675 & 29.00 & 8.01 & 10.2 & O\\ 
J1741$-$0840 & 2.0431 & 75.30 & 33/519 & 20.12 & 8.20 & 41.4 & -\\ 
J1741$-$2019 & 3.9045 & 74.90 & 306/1496 & 30.83 & 8.01 & 38.0 & O\\ 
J1741$-$3016 & 1.8937 & 382.00 & 7/4531 & 10.95 & 8.09 & 1.6 & -\\ 
J1741$-$3927 & 0.5122 & 158.29 & 1587/12300 & 20.83 & 8.01 & 14.7 & O\\ 
J1743$-$3150 & 2.4147 & 192.11 & 149/3479 & 20.93 & 8.01 & 11.0 & -\\ 
J1744$-$1610 & 1.7572 & 66.67 & 95/5350 & 19.92 & 8.02 & 18.1 & -\\ 
J1744$-$3130 & 1.0661 & 193.50 & 78/15760 & 16.13 & 8.02 & 4.0 & -\\ 
J1745$-$0129 & 1.0454 & 90.10 & 4/1522 & 9.74 & 8.22 & 1.4 & -\\ 
J1745$-$3040 & 0.3674 & 88.07 & 3516/11431 & 35.79 & 8.01 & 22.9 & O\\ 
J1747$-$2647 & 0.5003 & 570.00 & 1/20991 & 8.12 & 8.12 & 1.0 & -\\ 
J1748$-$1300 & 0.3941 & 99.52 & 22/3364 & 12.13 & 8.02 & 5.1 & -\\ 
J1750$-$3157 & 0.9104 & 205.98 & 14/6921 & 11.14 & 8.01 & 2.0 & -\\ 
J1751$-$3323 & 0.5482 & 295.61 & 3/15323 & 10.54 & 8.01 & 2.4 & -\\ 
J1751$-$4657 & 0.7424 & 20.60 & 234/1429 & 16.64 & 8.03 & 15.3 & G\\ 
J1752$-$2806 & 0.5626 & 50.32 & 8137/15146 & 37.22 & 8.01 & 58.8 & O\\ 
J1753$-$12 & 0.4055 & 73.00 & 1/3270 & 9.00 & 9.00 & 1.0 & -\\ 
J1753$-$38 & 0.6668 & 167.60 & 1/1193 & 10.98 & 10.98 & 1.0 & -\\ 
J1754$-$3510 & 0.3927 & 81.75 & 163/16044 & 15.52 & 8.01 & 3.0 & -\\ 
J1756$-$2225 & 0.4050 & 329.00 & 6/15557 & 16.43 & 8.01 & 2.9 & -\\ 
J1756$-$2435 & 0.6705 & 367.10 & 4/18794 & 8.76 & 8.06 & 1.5 & -\\ 
\hline
\end{tabular}
\end{center}
\end{scriptsize}
\end{table*}
\begin{table*}[ht]
    \renewcommand\thetable{\ref{table:properties}}
  \begin{scriptsize}
  \caption{Continued.}
  \renewcommand\arraystretch{1.0}
  \setlength{\tabcolsep}{2.0mm}  
  \begin{center}
  \begin{tabular}{@{\extracolsep{\fill}}cccccccc}
  \hline  
\multicolumn{1}{c}{Jname}  &    Period      &  DM        &     Number of Single     &     Max SP         &    Min SP  & Fluence  & Distribution         \\
 \multicolumn{1}{c}{}    &   (s)     &   (pc$\cdot$cm$^{-3}$)         &    Pulses Detected         &        S/N       &  S/N    & ratio &   class     \\
\hline
J1757$-$2223 & 0.1853 & 239.30 & 327/45334 & 27.75 & 8.01 & 16.1 & O\\ 
J1757$-$2421 & 0.2341 & 179.59 & 42/26913 & 11.31 & 8.02 & 3.6 & -\\ 
J1759$-$1956 & 2.8434 & 236.40 & 6/2215 & 11.61 & 8.52 & 3.5 & -\\ 
J1759$-$2205 & 0.4610 & 176.91 & 94/22779 & 14.48 & 8.05 & 4.8 & -\\ 
J1759$-$3107 & 1.0790 & 128.90 & 31/7786 & 10.30 & 8.05 & 2.6 & -\\ 
J1801$-$0357 & 0.9215 & 120.37 & 1/863 & 8.04 & 8.04 & 1.0 & -\\ 
J1801$-$2920 & 1.0819 & 125.61 & 233/5823 & 23.29 & 8.01 & 24.3 & L\\ 
J1802+0128 & 0.5543 & 98.00 & 1/3828 & 8.94 & 8.94 & 1.0 & -\\ 
J1803$-$2137 & 0.1337 & 233.94 & 317/65515 & 17.30 & 8.01 & 31.8 & O\\ 
J1803$-$2712 & 0.3344 & 165.50 & 1/19918 & 8.29 & 8.29 & 1.0 & -\\ 
J1805$-$1504 & 1.1813 & 225.00 & 3/5333 & 12.82 & 8.60 & 3.3 & -\\ 
J1806+1023 & 0.4843 & 52.00 & 38/3264 & 13.13 & 8.03 & 4.1 & -\\ 
J1807$-$0847 & 0.1637 & 112.38 & 981/30514 & 13.19 & 8.01 & 8.6 & G\\ 
J1807$-$2557 & 2.7642 & 385.00 & 1/1519 & 9.78 & 9.78 & 1.0 & -\\ 
J1808$-$0813 & 0.8760 & 151.27 & 1/8100 & 8.37 & 8.37 & 1.0 & -\\ 
J1808$-$2057 & 0.9184 & 606.80 & 20/13720 & 12.11 & 8.10 & 2.4 & -\\ 
J1808$-$3249 & 0.3649 & 147.37 & 49/29158 & 16.89 & 8.01 & 6.4 & -\\ 
J1809$-$2109 & 0.7024 & 381.63 & 28/17939 & 13.41 & 8.05 & 3.4 & -\\ 
J1810$-$5338 & 0.2611 & 45.00 & 32/3047 & 14.43 & 8.05 & 3.1 & -\\ 
J1812$-$1718 & 1.2054 & 251.40 & 3/10454 & 9.67 & 8.04 & 1.7 & -\\ 
J1814$-$0618 & 1.3779 & 168.00 & 1/2679 & 8.96 & 8.96 & 1.0 & -\\ 
J1815$-$1910 & 1.2499 & 547.80 & 1/13442 & 8.12 & 8.12 & 1.0 & -\\ 
J1817$-$3618 & 0.3870 & 94.30 & 26/2741 & 14.22 & 8.06 & 4.7 & -\\ 
J1817$-$3837 & 0.3845 & 102.85 & 5/2759 & 10.66 & 8.21 & 1.3 & -\\ 
J1818$-$1422 & 0.2915 & 619.65 & 36/28820 & 9.93 & 8.04 & 13.2 & -\\ 
J1819$-$1458 & 4.2632 & 196.00 & 21/1477 & 27.81 & 8.41 & 13.0 & -\\ 
J1820$-$0427 & 0.5981 & 84.44 & 659/3955 & 17.08 & 8.01 & 8.0 & L\\ 
J1820$-$0509 & 0.3373 & 102.11 & 9/12452 & 10.21 & 8.06 & 2.5 & -\\ 
J1822$-$2256 & 1.8743 & 121.20 & 470/5602 & 18.80 & 8.01 & 36.7 & L\\ 
J1823$-$0154 & 0.7598 & 135.75 & 1/5528 & 8.97 & 8.97 & 1.0 & -\\ 
J1823+0550 & 0.7529 & 66.73 & 24/1761 & 15.63 & 8.01 & 14.3 & -\\ 
J1823$-$1126 & 1.8465 & 607.00 & 33/2274 & 21.49 & 8.07 & 8.0 & -\\ 
J1823$-$3106 & 0.2841 & 50.25 & 12/3734 & 10.21 & 8.26 & 3.9 & -\\ 
J1824$-$0127 & 2.4995 & 62.93 & 2/1680 & 11.36 & 8.87 & 1.2 & -\\ 
J1824$-$1118 & 0.4358 & 601.31 & 1/14459 & 8.56 & 8.56 & 1.0 & -\\ 
J1824$-$1945 & 0.1893 & 224.27 & 4012/44369 & 19.76 & 8.01 & 28.1 & O\\ 
J1824$-$2233 & 1.1617 & 156.50 & 14/9039 & 11.18 & 8.14 & 1.8 & -\\ 
J1824$-$2328 & 1.5059 & 195.40 & 2/4184 & 8.98 & 8.23 & 3.7 & -\\ 
J1825$-$0935 & 0.7690 & 19.36 & 3376/10924 & 20.26 & 8.01 & 33.1 & O\\ 
J1825$-$1446 & 0.2792 & 352.23 & 786/30088 & 24.15 & 8.01 & 38.3 & O\\ 
J1825$-$33 & 1.2712 & 43.00 & 5/1251 & 16.72 & 10.02 & 5.7 & -\\ 
J1826$-$1334 & 0.1015 & 231.48 & 3/65018 & 9.64 & 8.69 & 1.4 & -\\ 
J1826$-$1419 & 0.7706 & 160.00 & 2/8176 & 11.33 & 8.50 & 1.7 & -\\ 
J1827$-$0750 & 0.2705 & 374.99 & 230/23292 & 16.72 & 8.03 & 19.0 & G\\ 
J1829$-$0734 & 0.3184 & 317.66 & 2/26384 & 9.72 & 8.96 & 1.1 & -\\ 
J1830$-$1059 & 0.4051 & 159.70 & 11/20738 & 11.53 & 8.42 & 2.0 & -\\ 
J1830$-$1135 & 6.2216 & 257.00 & 45/1687 & 17.90 & 8.01 & 6.7 & -\\ 
J1831$-$1223 & 2.8579 & 342.00 & 50/3674 & 21.63 & 8.05 & 38.3 & -\\ 
J1832$-$0827 & 0.6473 & 301.24 & 52/6488 & 13.67 & 8.01 & 4.5 & -\\ 
J1833$-$0827 & 0.0853 & 409.83 & 293/52766 & 17.96 & 8.01 & 17.8 & O\\ 
J1835$-$1020 & 0.3025 & 115.67 & 4/13887 & 9.31 & 8.21 & 1.1 & -\\ 
J1837$-$0045 & 0.6170 & 86.98 & 2/6807 & 11.42 & 8.58 & 1.0 & -\\ 
J1837$-$0653 & 1.9058 & 316.10 & 151/1102 & 22.68 & 8.01 & 18.1 & -\\ 
J1837$-$1243 & 1.8760 & 300.00 & 19/2239 & 12.93 & 8.12 & 2.4 & -\\ 
J1837$-$1837 & 0.6184 & 100.74 & 13/18320 & 10.86 & 8.33 & 2.2 & -\\ 
J1839$-$0141 & 0.9333 & 293.20 & 2/9001 & 9.09 & 8.08 & 1.7 & -\\ 
J1839$-$1238 & 1.9114 & 174.16 & 8/1098 & 10.81 & 8.06 & 1.9 & -\\ 
J1840$-$0626 & 1.8934 & 748.00 & 3/4437 & 12.75 & 8.13 & 1.4 & -\\ 
J1840$-$0809 & 0.9557 & 349.80 & 215/6592 & 15.15 & 8.02 & 9.8 & L\\ 
J1840$-$0815 & 1.0964 & 233.80 & 35/5746 & 17.61 & 8.01 & 3.2 & -\\ 
\hline
\end{tabular}
\end{center}
\end{scriptsize}
\end{table*}
\begin{table*}[ht]
    \renewcommand\thetable{\ref{table:properties}}
  \begin{scriptsize}
  \caption{Continued.}
  \renewcommand\arraystretch{1.0}
  \setlength{\tabcolsep}{2.0mm}  
  \begin{center}
  \begin{tabular}{@{\extracolsep{\fill}}cccccccc}
  \hline  
\multicolumn{1}{c}{Jname}  &    Period      &  DM        &     Number of Single     &     Max SP         &    Min SP  & Fluence  & Distribution         \\
 \multicolumn{1}{c}{}    &   (s)     &   (pc$\cdot$cm$^{-3}$)         &    Pulses Detected         &        S/N       &  S/N    & ratio &   class     \\
\hline
J1840$-$1419 & 6.5976 & 19.40 & 34/1273 & 17.85 & 8.19 & 9.1 & -\\ 
J1841$-$0157 & 0.6633 & 475.00 & 36/12664 & 10.20 & 8.04 & 3.9 & -\\ 
J1841$-$0310 & 1.6577 & 216.00 & 5/3800 & 10.64 & 8.76 & 1.6 & -\\ 
J1841+0912 & 0.3813 & 49.16 & 4/2086 & 9.28 & 8.02 & 1.6 & -\\ 
J1842+0257 & 3.0883 & 148.10 & 3/2040 & 9.44 & 8.30 & 1.5 & -\\ 
J1842$-$0359 & 1.8399 & 195.98 & 235/3424 & 20.48 & 8.01 & 48.6 & O\\ 
J1843$-$0000 & 0.8803 & 101.50 & 19/4771 & 15.10 & 8.03 & 3.6 & -\\ 
J1843$-$0211 & 2.0275 & 441.70 & 38/4143 & 12.58 & 8.14 & 5.3 & -\\ 
J1844+00 & 0.4605 & 345.50 & 2099/18242 & 24.30 & 8.01 & 29.2 & O\\ 
J1844$-$0433 & 0.9910 & 123.16 & 74/6357 & 19.77 & 8.01 & 7.2 & -\\ 
J1845$-$0434 & 0.4868 & 230.80 & 2/12944 & 8.72 & 8.42 & 1.3 & -\\ 
J1845$-$0826 & 0.6344 & 228.20 & 2/6621 & 9.75 & 8.06 & 1.3 & -\\ 
J1845$-$1351 & 2.6189 & 197.40 & 7/3207 & 15.90 & 8.50 & 2.3 & -\\ 
J1846$-$0257 & 4.4767 & 237.00 & 1/1407 & 8.95 & 8.95 & 1.0 & -\\ 
J1847$-$0402 & 0.5978 & 141.23 & 256/10539 & 14.54 & 8.01 & 5.6 & L\\ 
J1847$-$0605 & 0.7782 & 206.48 & 57/5397 & 16.73 & 8.02 & 3.5 & -\\ 
J1848$-$0123 & 0.6594 & 159.95 & 3908/9554 & 22.17 & 8.01 & 26.9 & O\\ 
J1848$-$1150 & 1.3122 & 163.40 & 16/4801 & 14.09 & 8.08 & 8.1 & -\\ 
J1848$-$1952 & 4.3082 & 18.23 & 21/800 & 22.55 & 8.21 & 26.4 & -\\ 
J1849$-$0636 & 1.4514 & 147.83 & 256/2894 & 30.38 & 8.01 & 17.5 & O\\ 
J1851$-$0053 & 1.4091 & 24.42 & 4/5962 & 10.97 & 8.14 & 4.3 & -\\ 
J1851+1259 & 1.2053 & 70.63 & 4/880 & 10.39 & 8.70 & 2.3 & -\\ 
J1852$-$0635 & 0.5242 & 173.58 & 162/8013 & 17.33 & 8.01 & 13.7 & -\\ 
J1854+0306 & 4.5578 & 192.40 & 1/1382 & 8.53 & 8.53 & 1.0 & -\\ 
J1854$-$1421 & 1.1466 & 130.40 & 1/693 & 11.69 & 11.69 & 1.0 & -\\ 
J1854$-$1557 & 3.4531 & 150.00 & 1/307 & 12.08 & 12.08 & 1.0 & -\\ 
J1857$-$1027 & 3.6872 & 108.90 & 82/500 & 26.97 & 8.03 & 17.0 & -\\ 
J1859+00 & 0.5596 & 420.00 & 19/15011 & 9.65 & 8.11 & 3.5 & -\\ 
J1900$-$2600 & 0.6122 & 38.16 & 60/1732 & 13.79 & 8.04 & 9.8 & -\\ 
J1901+0254 & 1.2997 & 185.00 & 55/4847 & 23.74 & 8.06 & 28.4 & -\\ 
J1901$-$0312 & 0.3557 & 106.40 & 57/23616 & 26.93 & 8.01 & 6.7 & -\\ 
J1901+0331 & 0.6555 & 401.24 & 1787/6408 & 25.25 & 8.01 & 25.6 & O\\ 
J1901$-$0906 & 1.7819 & 72.68 & 64/595 & 23.09 & 8.03 & 5.9 & -\\ 
J1901$-$1740 & 1.9569 & 24.40 & 53/1555 & 18.08 & 8.03 & 31.1 & -\\ 
J1902+0556 & 0.7466 & 177.49 & 16/11252 & 9.90 & 8.02 & 3.1 & -\\ 
J1903+0135 & 0.7293 & 244.96 & 2838/11518 & 30.17 & 8.01 & 25.2 & L\\ 
J1904+1011 & 1.8566 & 136.90 & 3/3393 & 11.88 & 8.96 & 1.3 & -\\ 
J1908+0457 & 0.8468 & 349.20 & 19/9920 & 11.49 & 8.14 & 2.1 & -\\ 
J1908+0500 & 0.2910 & 201.29 & 62/36083 & 14.53 & 8.01 & 2.8 & -\\ 
J1909+0007 & 1.0170 & 112.79 & 35/4130 & 11.52 & 8.03 & 3.2 & -\\ 
J1909+1102 & 0.2836 & 149.84 & 13/22213 & 9.51 & 8.01 & 1.9 & -\\ 
J1910+0358 & 2.3303 & 82.93 & 2/1802 & 9.53 & 8.08 & 1.9 & -\\ 
J1910+0728 & 0.3254 & 283.94 & 43/25815 & 14.53 & 8.06 & 2.1 & -\\ 
J1910+1231 & 1.4417 & 258.64 & 8/2913 & 10.57 & 8.65 & 2.0 & -\\ 
J1913$-$0440 & 0.8259 & 89.39 & 494/1927 & 21.18 & 8.01 & 9.3 & L\\ 
J1913+0446 & 1.6161 & 109.10 & 29/5198 & 24.18 & 8.08 & 5.8 & -\\ 
J1913+0904 & 0.1633 & 96.80 & 1/38594 & 8.03 & 8.03 & 1.0 & -\\ 
J1914+0219 & 0.4575 & 235.44 & 1/9180 & 8.13 & 8.13 & 1.0 & -\\ 
J1915+1009 & 0.4046 & 241.58 & 162/20765 & 14.50 & 8.01 & 3.8 & -\\ 
J1916+0748 & 0.5418 & 304.00 & 32/15506 & 11.61 & 8.07 & 9.7 & -\\ 
J1916+1023 & 1.6183 & 343.30 & 5/2595 & 10.25 & 8.29 & 7.4 & -\\ 
J1916+1312 & 0.2818 & 237.01 & 8/22355 & 8.97 & 8.05 & 1.9 & -\\ 
J1917+0834 & 2.1297 & 29.18 & 41/3944 & 14.51 & 8.04 & 6.1 & -\\ 
J1917+1353 & 0.1946 & 94.60 & 1/43161 & 9.19 & 9.19 & 1.0 & -\\ 
J1919+0021 & 1.2723 & 90.32 & 33/625 & 24.11 & 8.09 & 8.2 & -\\ 
J1919+0134 & 1.6040 & 191.90 & 28/2129 & 12.79 & 8.02 & 2.9 & -\\ 
J1920+1040 & 2.2158 & 308.80 & 3/3791 & 10.43 & 8.34 & 1.6 & -\\ 
J1938+0650 & 1.1216 & 70.80 & 6/709 & 10.36 & 8.02 & 2.1 & -\\ 
J1943+0609 & 0.4462 & 70.66 & 1/3543 & 8.79 & 8.79 & 1.0 & -\\ 
J1943$-$1237 & 0.9724 & 28.81 & 7/818 & 9.71 & 8.02 & 3.6 & -\\ 
\hline
\end{tabular}
\end{center}
\end{scriptsize}
\end{table*}
\begin{table*}[ht]
    \renewcommand\thetable{\ref{table:properties}}
  \begin{scriptsize}
  \caption{Continued.}
  \renewcommand\arraystretch{1.0}
  \setlength{\tabcolsep}{2.0mm}  
  \begin{center}
  \begin{tabular}{@{\extracolsep{\fill}}cccccccc}
  \hline  
\multicolumn{1}{c}{Jname}  &    Period      &  DM        &     Number of Single     &     Max SP         &    Min SP  & Fluence  & Distribution         \\
 \multicolumn{1}{c}{}    &   (s)     &   (pc$\cdot$cm$^{-3}$)         &    Pulses Detected         &        S/N       &  S/N    & ratio &   class     \\
\hline
J2005$-$0020 & 2.2797 & 35.93 & 1/465 & 9.06 & 9.06 & 1.0 & -\\ 
J2006$-$0807 & 0.5809 & 32.39 & 16/1369 & 12.90 & 8.02 & 7.8 & -\\ 
J2048$-$1616 & 1.9616 & 11.72 & 167/434 & 15.16 & 8.01 & 108.2 & -\\ 
\hline
\end{tabular}
\end{center}
\end{scriptsize}
\end{table*}



\end{document}